\documentclass[useAMS,usenatbib]{mnras}

\def\draftversion{2} 
\usepackage{amssymb,latexsym,graphicx,natbib,eufrak,times,amsmath,xspace,ifthen}

\ifthenelse{ \draftversion > 0 }{
} {
  
}

\ifthenelse{\equal{\draftversion}{1}}{
  \usepackage{xcolor}
  \newcommand{\sep}[1]{\par\begin{color}[rgb]{0,0.4,0}\begin{center}\hrule\end{center}\end{color}\par} 
  \newcommand{\todo}[1]{\begin{color}{red}\ \ifthenelse{\equal{#1}{}} {$\bullet\bullet\bullet$} {$\bullet$\ #1 $\bullet$}\end{color}} 
  \newcommand{\idea}[1]{\begin{color}[rgb]{0,0.4,0}\textit{#1}\end{color}} 
  \newcommand{\sk}[1]{\begin{color}[rgb]{0.6,0,0.6}#1\end{color}} 
  \newcommand{\toc}{\par\begin{color}[rgb]{0.6,0,0.6}\begin{center}\hrule\vspace{0.5mm}\begingroup\small\let\cleardoublepage\relax\let\clearpage\relax\mytoc\endgroup\vspace{0.5mm}\hrule\end{center}\end{color}\par} 
  \newcommand{\inprep}{\begin{color}[rgb]{0,0.8,0.8}in preparation\end{color}\xspace}
  
  }{
  \newsavebox{\trashcan}

  \newcommand{\sep}[1]{}
  \newcommand{\todo}[1]{}
  \newcommand{\idea}[1]{}
  \newcommand{\sk}[1]{}
  \newcommand{\toc}{}
  \newcommand{\inprep}{in preparation\xspace}
  
  }
 \makeatletter\newcommand\mytoc{\@starttoc{toc}}\makeatother 
\long\def\symbolfootnote[#1]#2{\begingroup%
\def\thefootnote{\fnsymbol{footnote}}\footnote[#1]{#2}\endgroup} 

\newcommand{\eqn}[2][]{Equation#1~\ref{eqn:#2}} 
\newcommand{\fig}[2][]{Figure#1~\ref{fig:#2}}
\newcommand{\tab}[2][]{Table#1~\ref{tab:#2}}
\newcommand{\sect}[2][]{Section#1~\ref{sec:#2}}
\newcommand{\app}[2][]{Appendix#1~\ref{sec:#2}}

\newcommand{\bb}[1]{\ifmmode \mbox{\boldmath $ #1$} \else  \mbox{\boldmath $#1$} \fi}

\newcommand{\mh}{\ensuremath{\textrm{\,--\,}}}    
\newcommand{\U}[1]{\ensuremath{\mathrm{~#1}}}     
\newcommand{\e}[1]{\ensuremath{\times 10^{#1}}}   

\newcommand{\yr}{\U{yr}}
\newcommand{\Myr}{\U{Myr}}          
\newcommand{\Gyr}{\U{Gyr}}          
\newcommand{\pc}{\U{pc}}
\newcommand{\kpc}{\U{kpc}}
          
\newcommand{\Msun}{\U{M}_{\odot}}   \newcommand{\msun}{\Msun}
\newcommand{\Msunyr}{\Msun\yr^{-1}} \newcommand{\msunyr}{\Msunyr}
   
\newcommand{\cc}{\U{cm^{-3}}}
\newcommand{\K}{\U{K}}
    
\newcommand{\kms}{\U{km\ s^{-1}}}

\newcommand{\hi}{H{\sc i}\xspace}                       
\newcommand{\htwo}{\ensuremath{\mathrm{H}_2}\xspace}                          
\newcommand{\tdep}{\ensuremath{t_\mathrm{dep}}\xspace}        

\newcommand{\ramses}{{\small RAMSES}\xspace}

\newcommand{\magi}{{\small MAGI}\xspace}

\newcommand{\vintergatan}{{\small VINTERGATAN}\xspace}

\newcommand{\lund}{Department of Astronomy and Theoretical Physics, Lund Observatory, Box 43, SE-221 00 Lund, Sweden}

\newcommand{\chalmers}{Department of Space, Earth and Environment, Chalmers University of Technology, SE-41296 Gothenburg, Sweden}

\newcommand{\fgas}{\ensuremath{f_{\rm gas}}\xspace}
\newcommand{\qrf}{\ensuremath{\mathcal{Q}_{\rm RF}}\xspace}
\newcommand{\lrf}{\ensuremath{\lambda_{\rm RF}}\xspace}

\newcommand{\gs}{\ensuremath{\mathcal{F}10}\xspace}
\newcommand{\gm}{\ensuremath{\mathcal{F}25}\xspace}
\newcommand{\gl}{\ensuremath{\mathcal{F}40}\xspace}

\title[Stability of galactic discs]{From giant clumps to clouds I:\\the impact of gas fraction evolution on the stability of galactic discs}
\author[Renaud, Romeo \& Agertz] {Florent~Renaud$^1$\thanks{florent@astro.lu.se}, Alessandro~B.~Romeo$^2$, Oscar~Agertz$^1$\\
$^1$ \lund\\
$^2$ \chalmers}

\date{Accepted 2021 September 09. Received 2021 September 07; in original form 2021 May 31}

\begin{document}
\maketitle


\begin{abstract}
The morphology of gas-rich disc galaxies at redshift $\sim 1\mh 3$ is dominated by a few massive clumps. The process of formation or assembly of these clumps and their relation to molecular clouds in contemporary spiral galaxies are still unknown. Using simulations of isolated disc galaxies, we study how the structure of the interstellar medium and the stability regime of the discs change when varying the gas fraction. In all galaxies, the stellar component is the main driver of instabilities. However, the molecular gas plays a non-negligible role in the inter-clump medium of gas-rich cases, and thus in the assembly of the massive clumps. At scales smaller than a few 100 pc, the Toomre-like disc instabilities are replaced by another regime, especially in the gas-rich galaxies. We find that galaxies at low gas fraction (10\%) stand apart from discs with more gas, which all share similar properties in virtually all aspects we explore. For gas fractions below $\approx 20\%$, the clump-scale regime of instabilities disappears, leaving only the large-scale disc-driven regime. Associating the change of gas fraction to the cosmic evolution of galaxies, this transition marks the end of the clumpy phase of disc galaxies, and allows for the onset of spiral structures, as commonly found in the local Universe.
\end{abstract}

\section{Introduction}

Over the last decade, developments in the modeling of galactic physics have significantly improved the realism of galaxy simulations. Models of stellar feedback in particular allow to successfully reproduce a number of observables of the overall structure of galaxies in cosmological context \citep{Naab2017}, and also in the properties of their interstellar medium and star forming clouds in isolated models \citep{Grisdale2017, Grisdale2018, Grisdale2019}. This is accompanied by formalisms and tools to analyze and characterize the structures of discs and their regimes of instabilities \citep[e.g.][]{Romeo2010, Elmegreen2010, Agertz2015b}. These works have mostly focused on contemporary disc galaxies, while their high redshift counterparts have often been neglected due to limited observational constraints on resolved star forming regions at $z\gtrsim 1$. However, gravitational lensing \citep{Cava2018, Dessauges2019} and forthcoming instruments like JWST open new perspectives.

At redshift $z\approx 1\mh 3$, the morphology of disc galaxies is often dominated by discrete, massive star forming clumps ($\sim 0.5 \mh 1 \kpc$, $\sim 10^8 \Msun$, \citealt{Cowie1995, Elmegreen2007, Elmegreen2009, Wuyts2012, Swinbank2015, Zanella2015, Fisher2017, Guo2018}). Surveys of high redshift galaxies show the dependence of the fraction of clumpy galaxies on the stellar mass and redshift, with more than half of the disc galaxies at $z\sim 3$ exhibiting clumpy morphologies \citep{Guo2015}. As the clumpy phase of galaxy evolution occurs at ``cosmic noon'', i.e. the peak of the cosmic star formation rate (SFR, \citealt{Madau2014}), it is of paramount importance in the assembly of the stellar populations observed in contemporary galaxies. Massive clumps have been invoked to participate in radial inflows and the assembly of the bulge \citep{Noguchi1999, Ceverino2010, Bournaud2014}, in the bimodal distribution of stellar abundances in the Milky Way \citep{Clarke2019, Khoperskov2021, Renaud2021}, and even in the onset of kinematically thick discs (van Donkelaar et al., \inprep). Yet, fundamental properties of these clumps, like their lifetimes, are still debated (\citealt{Genel2012, Mandelker2014, Mandelker2017, Oklopcic2017}, see a discussion in \citealt{Fensch2021}). One key problem is to explain when, how and how fast the transition from clumpy structures to spiral arms occurs along the course of galaxy evolution, and how this relates to mergers and to the gas accretion rate of galaxies \citep{Bournaud2007, Agertz2009b, Dekel2009, Ceverino2012, Cacciato2012}.

In the ``bottom-up'' scenario, massive clumps assemble by mergers of smaller structures \citep{Behrendt2016, Behrendt2019, Benincasa2019}. The alternative ``top-down'' hypothesis invokes the fragmentation of large structures into smaller sub-clumps \citep[see a discussion in][]{Behrendt2019}. Observations of high redshift clumpy galaxies have not been able to tell the two scenarios apart, as they suffer from incompleteness in the detection of clumps \citep{Tamburello2017, Faure2021}. Furthermore dust has been suggested to bias the interpretation of the hierarchy of massive clumps \citep{Buck2017}. From a theoretical perspective, in order to disentangle the two options and establish which dominates under different physical conditions, it is necessary to assess the stability regimes of disc galaxies, and their evolution with redshift. \citet{Inoue2016} conducted such an analysis using a suite of cosmological zoom simulations. By tracing the origins of massive clumps back to the pre-collapse protoclump region of the disc, they reported that, counter-intuitively, the material which eventually assembles into clumps lies in gravitationally stable regions, as estimated with a two-component stability criterion. To explain this contradiction, they invoked non-linear external mechanisms like mergers and tidal compression. Yet, a number of simulations of discs without a cosmological context successfully reproduce properties of clumpy galaxies \citep[e.g.][]{Bournaud2014, Perret2014, Fensch2017, Clarke2019}.

The lack of a clear understanding of the nature and the formation process of massive clumps hinders the field of galaxy evolution, and prevents us from connecting the physics of gas-rich galaxies to that of contemporary discs, which is much better understood. The natural approach to address these questions is to use cosmological simulations of disc galaxies from high to low redshift. However, this suffers from two major drawbacks. First, the numerical cost of resolving star forming clouds/clumps in cosmological simulations is still prohibitive and, at best, allows studying only a single or small number of cases. Second, by accounting for many mechanisms simultaneously active (gas accretion, mergers, intrinsic evolution), it is notoriously difficult to identify the physical role of a single parameter. For instance, in the cosmological zoom simulation \vintergatan, we found that the end of star formation in the thick disc coincides with both the epoch of the last major merger, and the end of the clumpy phase \citep{Agertz2021,Renaud2021,Renaud2021b}, thus blurring the interpretation of the exact role of each mechanism.

As a complement to cosmological zoom simulations, in this paper we present results from controlled numerical experiments of several disc galaxies, where the only parameter changed is the gas mass fraction. This mimics the cosmological evolution of the gas fraction in disc galaxies, as observed from $z \sim 1\mh 2$ to $z=0$ \citep{Saintonge2013}. We introduce our simulation method in section \sect{method}. We analyze the structural properties of these discs (\sect{structure}), and the physics of their instabilities (\sect{stability}). For the first time, we consider the mutual effects of the 3 components (atomic gas, molecular gas and stars) in the stability analysis, by applying the multi-component formalism of \citet{Romeo2013} to simulated galaxies. We highlight the existence of two regimes of instabilities in gas-rich discs, with a transition scale between the two at a few $100 \pc$ (\sect{scale}).

\section{Method}
\label{sec:method}

We simulate disc galaxies in isolation, i.e. without cosmological context, with the adaptive mesh refinement code \ramses \citep{Teyssier2002}. Apart from the cosmological aspects, the numerical method is identical to that of the \vintergatan simulation, i.e. it accounts for heating from a cosmological background \citep{Haardt1996}, calibrated at redshift 0 for all simulations for simplicity, atomic and molecular cooling down to $10 \K$, star formation at a fixed efficiency of 10\% per free-fall time above a density threshold of 100 \cc, and stellar feedback including winds, radiation pressure, type-II and Ia supernovae, and the production of oxygen and iron (see \citealt{Agertz2021} and references therein for details). The baryonic mass threshold to refine the grid cells is $\approx 1700 \Msun$. The resolution in the densest regions reaches $12 \pc$. All our analysis is conducted with this numerical resolution, but some quantities (like velocity dispersions) are measured using statistical tools over larger scales (``beam sizes'') of $70$ and $700 \pc$.

For each simulation, a galaxy of total mass $7\e{11} \msun$ is initialized with the \magi code \citep{Miki2018} with a dark matter halo, a central bulge, a stellar and a gaseous disc. The mass resolution of the initial conditions is $1.3\e{5} \Msun$ for the dark matter, and $4.6\e{4} \Msun$ for the stars. The initial parameters are provided in \app{init}. The galaxy is placed at the center of a $200 \kpc$ cube, filled with an ambient medium of $10^{-5} \cc$. The halos are truncated at a radius of $50 \kpc$, well-beyond the radial range considered in this paper, such that our analysis of the disc stability in unaffected by this artifact. All models are strictly identical, with total disc mass of $5.5\e{10} \Msun$, a bulge mass of $0.3\e{10}\Msun$. The only difference between the models is the relative mass of the stellar component and the gaseous disc, controlled by the gas fraction defined as
\begin{equation}
\fgas = \frac{M_{\rm gas}}{M_{\star} + M_{\rm gas}},
\end{equation}
where $M_{\star}$ sums the masses of the stellar disc and the bulge. (The bulge to total ratio remains the same in all models.)

We setup 3 models, with initial gas fractions of 15\%, 50\% and 70\%. Because of star formation and outflows, and the lack of cosmological accretion, this fraction naturally decreases with time. After a relaxation phase of $\sim 100 \mh 150 \Myr$, instabilities caused by the artificial initial conditions have vanished, and the star formation rate (SFR) stabilizes at $\approx 10$, 80 and $130 \msunyr$, respectively. This epoch is used for all our analyses. The gas fractions in the discs (i.e. in a slab covering $\pm 2.5 \kpc$ above and below the mid-plane of the disc, i.e. 5 times the scale height of the disc set in the initial conditions) at the epoch of analysis are approximately $\fgas = 10\%$, $25\%$ and $40\%$, and are used below to name the models: \gs, \gm, \gl. When measured in a sphere of $10 \kpc$ radius, i.e. including coronal gas and stellar spheroid, these fractions are $15\%$, $30\%$ and $50\%$. The stars formed out of gas during the simulation (as opposed to stars in the initial conditions) represent 3, 30 and 50\% of the stellar content of the discs, respectively.

To evaluate the fraction of atomic (\hi) and molecular hydrogen (\htwo), we use the cooling tables from \ramses under the assumptions of collisional ionization equilibrium in each individual cell to determine the ionization level. Then, in the neutral medium, the \htwo fraction is computed using the \citet{Krumholz2009} formalism. We then measure molecular gas to stellar mass ratios of 6\%, 15\% and 30\% in our three simulations in the snapshots considered.

These simulations are not tailored to depict three evolutionary epochs of the same galaxy. For instance, the disc of a given galaxy is expected to grow radially between the epochs where its gas fraction is 40\% and 10\% \citep{Kretschmer2020, Agertz2021}. Instead, the models presented here are rather part of a controlled experiment, where as few parameters as possible are changed between runs (in our case, only the initial gas fraction). For the same reason, we chose to consider 3 different simulations rather than a unique one studied at different epochs. Our simulations are therefore not directly probing an evolutionary effect, but rather the influence of a given parameter, which happens to evolve with redshift along galaxy evolution.

\section{Disc structure}
\label{sec:structure}

\subsection{Morphology}
\label{sec:morphologies}

\begin{figure*}
\centering
\includegraphics{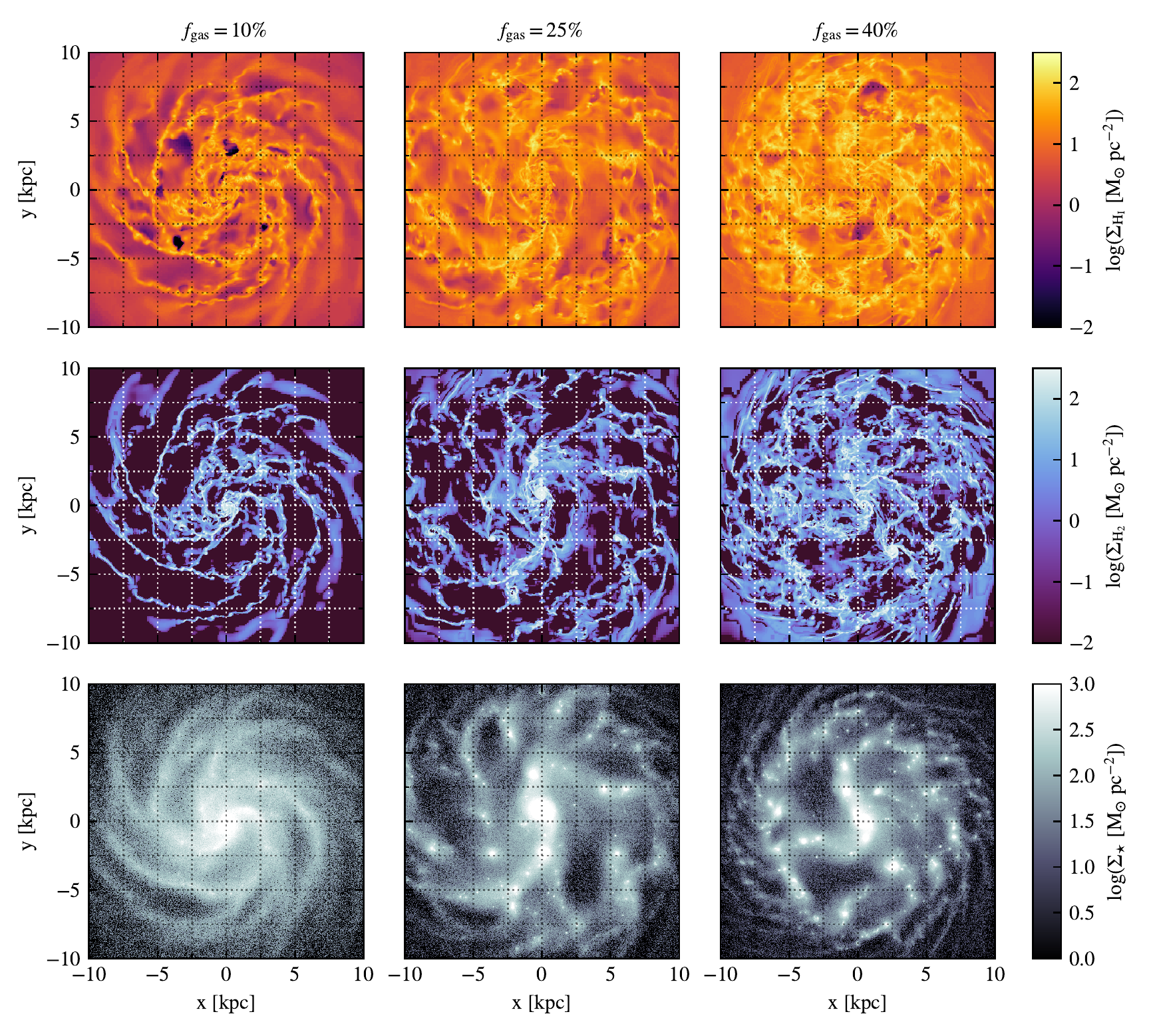}
\caption{Surface density maps of atomic gas (top), molecular gas (middle) and stars (bottom) of the 3 galaxies, at the epoch of the analysis. The gas fraction influences the size and relative importance of clumps and arms.}
\label{fig:maps}
\end{figure*}

The gas and stellar density maps of the simulations at the epochs analyzed are shown in \fig{maps}. The different gas fractions lead to different disc structure, at both small and large scales. In \gs, a grand-design spiral structure appears beyond the central kpc. However, spiral arms are not in place at higher gas fractions (\gm, \gl): the morphology there is rather dominated by less-organized gaseous and stellar clumps spanning a large range of sizes, and at no particular location within the disc. This resembles the morphology of clumpy discs, observed to be gas-rich galaxies at $z\sim 1\mh 3$ \citep[e.g.][]{Adamo2013}.

We note that all our discs contain cloud/clumps connected by elongated features. The main difference between the gas fractions is the relative importance of the two types of structures (spirals/elongated arms vs. clouds/clumps). At low gas fraction, the elongated arms are clearly identifiable as spiral arms, and connect virtually all over-densities, as so-called beads on a string \citep{Elmegreen1983, Elmegreen2019, Renaud2013b, Elmegreen2018}. The sizes of these clouds and stellar clumps appear to be capped by the width of the spiral arms, and thus set by a large-scale instability process. However, in the two gas-rich cases, the elongated features are sub-dominant with respect to massive, yet local, structures made of gaseous and stellar clumps. There, the sizes of clumps appears to be independent of their surroundings, suggesting that a local, sub-kpc process of instability plays a bigger role.

Contrarily to the spiral arms in \gs, the elongated structures have less-coherent orientations and lengths across the \gm and \gl discs. Their density profiles are also much shallower than that of the thinner spiral arms. This is largely due to the presence of an intermediate density medium in the inter-clump areas, while the inter-spiral arm medium of \gs is significantly more diffuse and even absent in the molecular phase. A close examination of the morphology of the elongated features in \gm and \gl reveals that some of them, but not all, connect to clumps in the S-shape characteristic of tidal tails. Some of these tails form indeed from clump-clump interactions in the two gas-rich cases, which also participates in populating the inter-clump medium with intermediate density gas ($\sim 100 \Msun\,\pc^{-2}$ in both the atomic and molecular phases). This situation does not occur in \gs where the clouds follow the organized motion of their spiral arm hosts, and very rarely interact.

\subsection{Scale-dependence of the gas structures}
\label{sec:psd}

\begin{figure}
\centering
\includegraphics{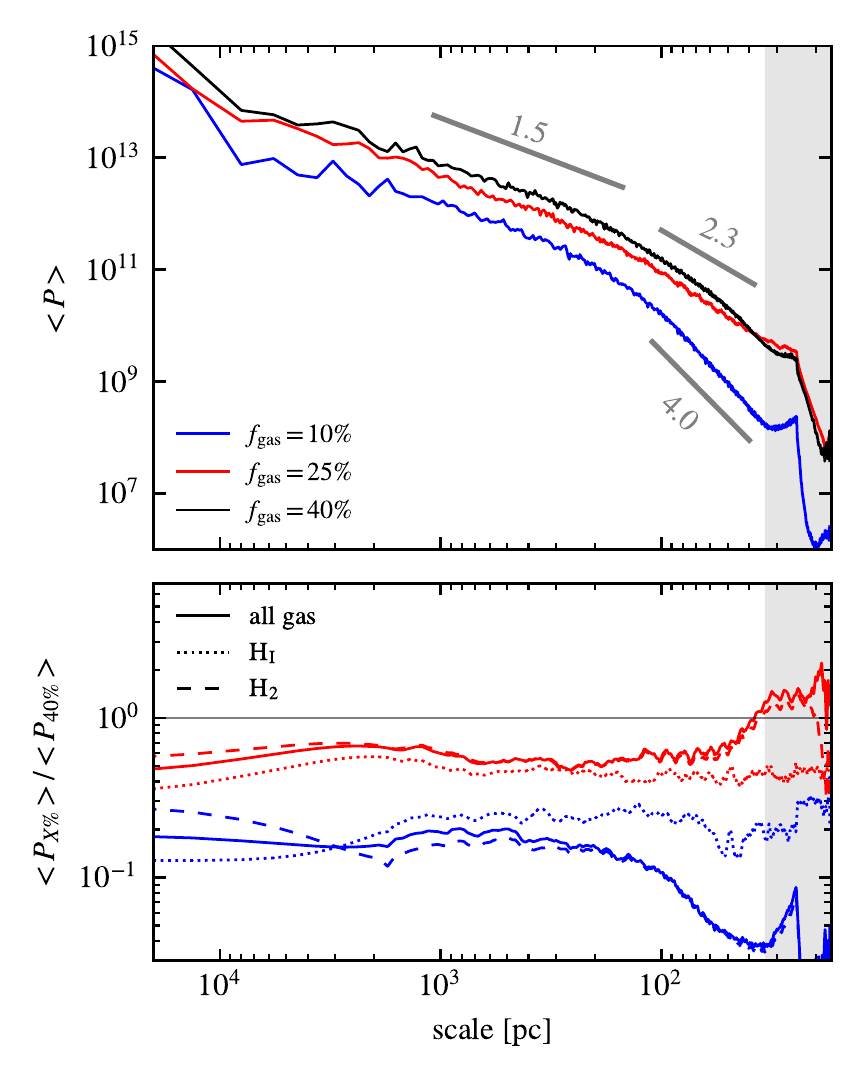}
\caption{Top: PSD of the two-dimensional maps of total gas surface densities. Several power-laws $\langle P\rangle \propto l^{\alpha}$ and the corresponding indices are indicated. Bottom: PSDs of all the gas, of \hi and of H$_2$ in galaxies with $\fgas = 10\%$ (\gs, blue) and $25 \%$ (\gm, red), normalized by those from the case $\fgas = 40\%$ (\gl). Differences between the galaxies mostly appear at small scales (in the three-dimensional regime of turbulence), and are caused by the response of the molecular phase to different instability regimes.}
\label{fig:psd}
\end{figure}

To quantify the differences in the organization of the ISM structure spotted by eye in the previous section, the top panel of \fig{psd} shows the power spectrum densities (PSDs) of the face-on gas surface density (atomic + molecular). The data in the gray-shaded area is likely affected by sampling issues due to our limited resolution, and is thus ignored in the interpretation. On the other hand, the largest scales shown here are close to the total size of the simulation volume, and thus suffer from under-sampling.

At large and intermediate scales (from $\approx 150 \pc$ to $\approx 6 \kpc$), the three galaxies yield a power-law regime $\langle P\rangle \propto l^{\alpha}$ with $\alpha \approx 1.5$, where $l$ is the spatial scale. This is compatible, yet a bit shallower, with the power-spectra of \hi in the THINGS galaxies ($\approx 1.5\mh 2.5$, \citealt{Walter2008, Walker2014, Grisdale2017, Dib2021}). While the different normalizations of the PSDs correspond to the different total gas masses, the similarity of the slopes for the 3 galaxies indicates that the organization of the ISM at these large and intermediate scales is largely independent of the gas fraction. Our results thus suggest that the variations of power-spectra observed in real galaxies (notably at scales $\gtrsim 1 \kpc$) are caused by other factors, possibly the mass and/or the size of the discs, and the inclination (see \citealt{Dib2021}, on the effects of inclination on the power spectra of \hi in simulated galaxies).

At $\approx 100 \mh 150 \pc$, the PSDs transition into steeper regimes. Previous works showed that this transition corresponds to the scale-height of the disc, below which turbulence can develop inside the disc (i.e. at scales smaller than the disc scale-height) along the vertical dimension, in addition to the two-dimensional modes in the plane of the disc (\citealt{Elmegreen2001, Padoan2001, Dutta2009, Renaud2013b}). Since such turbulence has an imprint on the three-dimensional structure of the ISM, its signature is also visible in projected maps, as shown here. At the epochs analyzed, the gas component of our galaxies yields exponential scale-heights of 92, 145 and $170 \pc$ respectively, corresponding to the transition visible in the PSDs, and thus confirming this interpretation.

At scales smaller than the transition ($\lesssim 100\mh 150 \pc$), the PSDs of the three galaxies diverge: \gs reaches a steep $\alpha \approx 4.0$, while the intermediate case (\gm) follows $\alpha \approx 2.3$. The power spectrum of the most gas-rich case (\gl) continuously gets steeper with decreasing scale, without converging toward a fixed value. It is however possible that it would converge at smaller scales, not accessible in our simulations. This divergence in slope indicates that the transition from large scales to small scales is not only caused by probing the three-dimensional nature of turbulence, but that other mechanisms must be invoked at scales below $\sim 100 \pc$. They also show that these mechanisms are scale-dependent, and vary with the gas fraction.

To gather more clues on the origins of these differences, the bottom panel of \fig{psd} shows the PSDs of all the gas, of \hi only, and of \htwo only, for the \gs and \gm cases, each normalized by their equivalent in the \gl simulation. We first note that the relatively flat dotted curves indicate that the structure of \hi is comparable at all gas fractions, even at small scales. This phase, found typically at low-density, includes the outer envelopes of star forming clouds, as well as the inter-cloud/arm regions.

Conversely, the PSDs of the \htwo component differ significantly between our cases, and carry most of the differences in the PSDs at small scales seen in the top panel ($\lesssim 100 \mh 150 \pc$). Therefore, the differences in the organization of the ISM occur mainly in the molecular phase, and is partly connected to the three-dimensional regime of turbulence within the disc scale-height. This corresponds to the visual impression of different morphologies between the molecular clouds at low gas fraction and the giant clumps for higher \fgas, and thus of differences in the structure of the star forming material, and ultimately in the clustering of star formation and feedback.

The PSDs of the different phases highlight similarities between the two highest gas fractions (\gm and \gl), while \gs is significantly different. This hints to a transition regime in the structure of the ISM, particularly in the molecular phase, for gas fractions $\approx 20\%$ (see van Donkelaar et al. \inprep, for another measurement of this value). Additional simulations would be required to establish the details of such a transition. We suspect, however, that such details depend on other parameters, like galactic mass and/or morphology. Such an extensive study is beyond the scope of this paper.

Apart from different cooling processes, there is a priori no reason for the molecular nature of the small-scale structures to be responsible for the organization of the ISM. Rather, the accumulation of dense and cold gas in large and deep potential wells allows for the assembly of dense, cold and partly self-shielded clouds and clumps, in which the gas becomes molecular. In other words, the variations in the molecular content of discs is more likely a consequence than a cause of the overall structure of the ISM. In this case, the differences in the molecular gas structure should be detectable as different instability regimes, varying with the gas fraction, i.e. a different imbalance between turbulent pressure support, rotation, and gravity. The results from this section also suggest different instability regimes between large and small scales. We explore and validate these hypotheses in the \sect[s]{stability} and \ref{sec:ab}.

\subsection{Gas density distributions and the origins of high SFRs}

\begin{figure}
\centering
\includegraphics{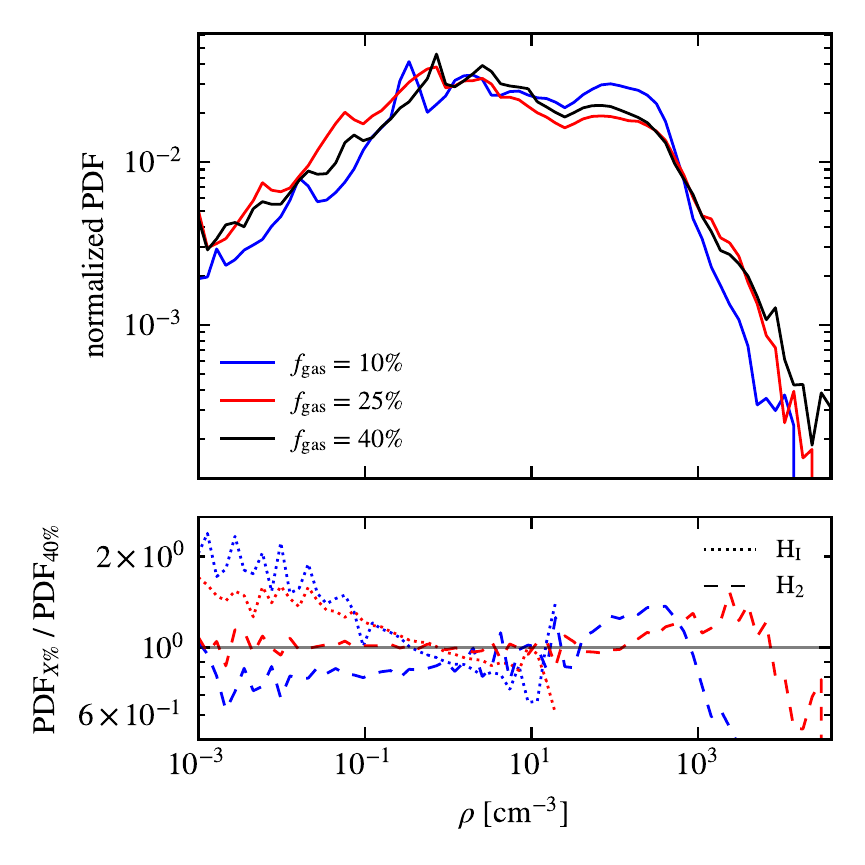}
\caption{Top: mass-weighted gas density PDFs for the 3 gas fractions considered, normalized by the total gas mass. Bottom: normalized PDFs of the \hi and \htwo components only, normalized by their equivalent in the \gl case. The gas-rich cases yield more dense gas, but not particularly denser gas.}
\label{fig:pdf}
\end{figure}

The top-panel of \fig{pdf} shows the probability distribution functions (PDFs) of the volume density of gas in our simulations. All three cases approximately follow a log-normal functional form, as expected from turbulence theory \citep{Hennebelle2012}. The main differences between the galaxies are a slight excess of intermediate density gas ($\approx 10^2 \cc$) in the low gas-fraction case, corresponding to spiral arms, and the maximum density increasing with the gas fraction, but only by a factor of $\approx 2$ between the extreme cases. By examining the ratios of the PDFs in \hi and \htwo (bottom panel), we find again similarities between the \gm and \gl, specially in \htwo up to the highest densities, i.e. in the clumps themselves. More important differences are found with the lowest gas fraction. In particular, we note a deficit of diffuse \hi in the gas-rich galaxies, with differences amplifying with decreasing density. This corresponds to the efficient launch of galactic outflows due to the more vigorous SFR in the gas-rich cases compared to \gs.

In \htwo, the PDFs are similar up to $\approx 400 \cc$, but in the denser media, the two gas-rich galaxies yield a significant excess of dense molecular gas with respect to the $\fgas = 10\%$ case. Such a medium is found in the central-most regions of the individual clouds and clumps. This excess is due to more numerous dense structures, and also significantly larger volumes of dense gas in the massive clumps of the gas-rich galaxies, as suggested by the power spectra (\fig{psd}). For instance, the molecular phase denser than $10^3 \cc$ spans regions typically $\approx 100 \pc$ wide in \gl, but only $30 \pc$ in \gs, thus corresponding to a factor $\approx 40$ smaller in volume. Therefore, the molecular structures in gas-rich galaxies are not significantly denser than in galaxies with lower gas fractions, but span larger volumes, thus corresponding to larger masses.

Between our most extreme simulations, the SFR varies by a factor of 13 but this is balanced when normalizing by the gas mass: the depletion time ($\tdep = M_{\rm gas} / {\rm SFR}$) changes by only a factor of 3. Therefore, the gas fraction of disc galaxies has a much stronger influence on the SFR than on the depletion time. In other words, the higher SFR of the gas-rich galaxies, compared to the case at low gas fraction, originates from larger amount of star forming gas, rather than a more efficient (or more rapid) star formation. Despite the SFRs of our gas-rich examples (respectively 80 and $130 \msunyr$ in \gm and \gl) being similar to that at the peaks of starbursts in local galaxy mergers (e.g. $\sim 100 \msunyr$ in simulated major mergers of galaxies in the mass range considered here, \citealt{Renaud2014b}), the shapes of their PDFs are radically different. In interacting galaxies, a starburst activity correlates with the presence of an excess of dense gas, i.e. larger amounts of dense gas \emph{and} higher densities \citep{Renaud2019b}. This excess of dense gas implies reduced depletion times, i.e. fast star formation, typically $\sim 10^{-2} \mh 10^{-3} \Gyr$ in the local Universe for the galactic masses considered here (e.g. \citealt{Martinez2012, Diaz2020}). In our isolated discs however, the galactic-wide depletion times are $\approx 0.6 \Gyr$, $0.3 \Gyr$ and $0.2 \Gyr$ in \gs, \gm and \gl respectively, i.e. comparable to that measured in isolated systems at low and high redshift ($\sim 0.1\mh 1 \Gyr$, see e.g. \citealt{Bigiel2010, Saintonge2011, Leroy2013, Tacconi2013}). Therefore, similar SFRs are reached in isolated gas-rich discs by forming \emph{many} stars, and in local starbursting mergers by forming stars \emph{fast}. This confirms that different physical mechanisms are active and/or that a turbulence of different nature is present at cloud/clump scales in different galaxy types, even with similar SFRs \citep{Renaud2014b}. The star formation prescription adopted in our models (i.e. a fixed efficiency of a few percents per free-fall time) is supported by observations of nearby clouds in the Milky Way \citep[e.g.][]{Elmegreen2002, Krumholz2007}, but could well be different in extreme environments like turbulent gas-rich discs. This would then alter the details of the star formation activity in the massive clumps. However, our conclusions remain qualitatively valid for any star formation relation with a super-linear dependence on the density, as widely accepted.

\section{Disc instabilities}
\label{sec:stability}

Since our three galaxies are setup with the same disc-bulge-halo combination and masses, the morphological and structural differences noted above result from the different relative importance of the dissipative component (gas) on the stability regimes of the discs. This section presents how instabilities arise.

\subsection{Radial profiles and the failure of single-component stability criteria}
\label{sec:radial}

\begin{figure*}
\centering
\includegraphics{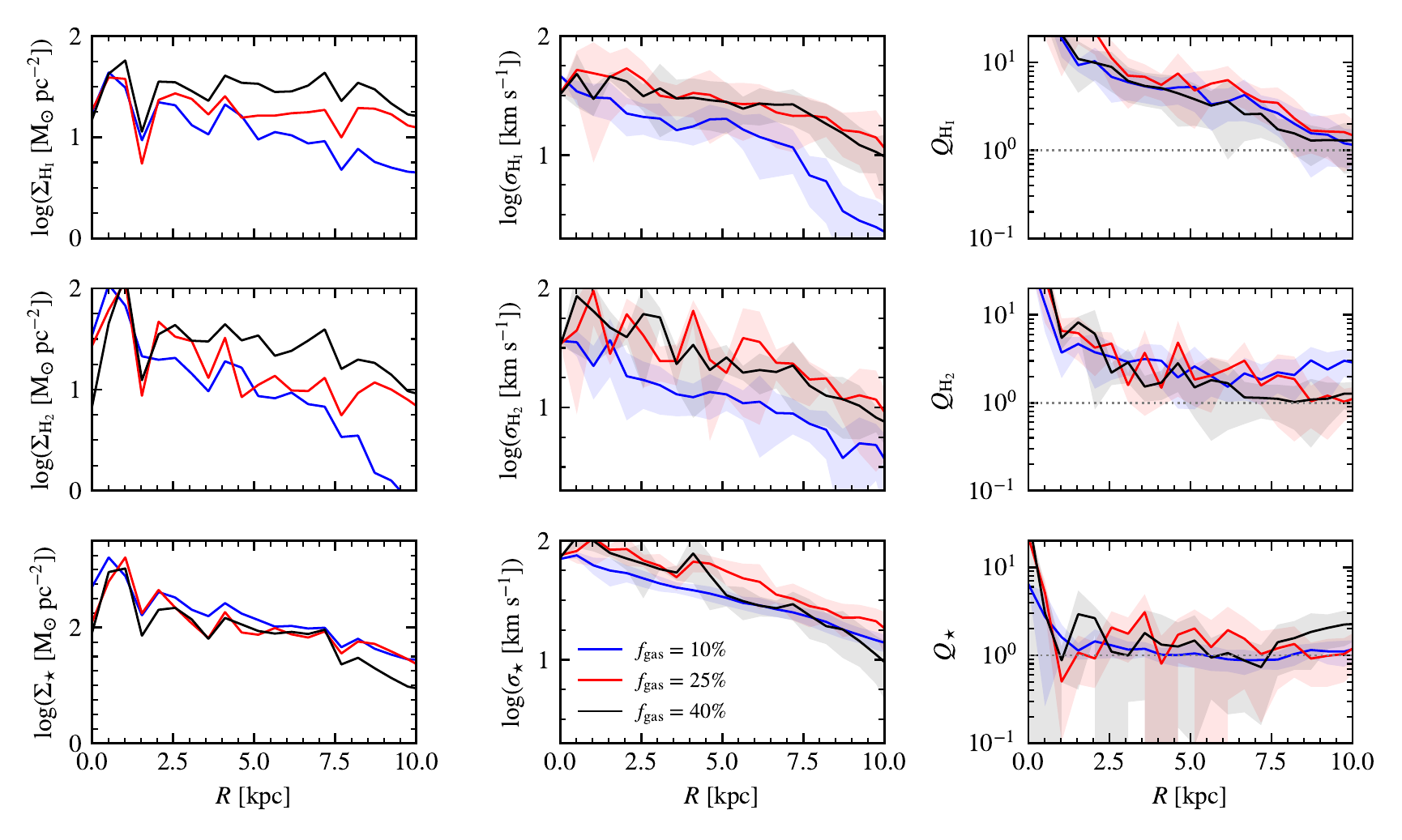}
\caption{Radial profiles of the surface densities (left), radial velocity dispersion (center), and $Q_i$ stability parameter (\eqn{qt}, right), for the atomic gas (top), molecular gas (middle), and stars (bottom). All quantities are measured on a scale of $700\pc$, and radially binned at a scale of $1 \kpc$. The curves show the median value in each radial bin, and the shaded areas indicate the 1-$\sigma$ robust dispersion of the velocity dispersion.}
\label{fig:profilelowres}
\end{figure*}

Under the approximation of an infinitesimally thin disc, the stability parameter for the individual component $i$ (gas or stars) reads
\begin{equation}
\label{eqn:qt}
Q_i = \frac{\kappa \sigma_i}{\pi G \Sigma_i},
\end{equation}
where $\kappa$ is the epicyclic frequency, $\sigma_i$ the radial velocity dispersion, and $\Sigma_i$ the surface density (see \citealt{Safronov1960, Goldreich1965} on the parameter for fluids, and \citealt{Toomre1964} on that for stars\footnote{For the stellar component, one must apply the collisionless version of the Boltzmann equation (sometimes referred to as the Vlasov equation, but see \citealt{Henon1982} for historical considerations) to the phase-space distribution function. In this case, the factor $\pi$ is replaced with the numerical value $\approx 3.36$ coming from the derivation of a modified Bessel function. In practice, the difference on the stability regime between using $\pi$ and 3.36 is negligible (see \citealt{Binney2008}, their Figure 6.13). We thus use the same numerical factor for all components.}). These three terms represent the rotational support of the disc, the pressure force, and the self-gravity of the $i$-th component, respectively. Under this formalism, discs are traditionally considered unstable where $Q_i \lesssim 1$. 

To apply this formalism to our models, we adopt the resolution of $700 \pc$ to post-process our simulations\footnote{The numerical resolution of the simulations remains $12 \pc$ in the entire paper: only the scale of measurement of quantities is changed in post-processing.}. The motivation for this scale is two-fold: (\emph{i}) it corresponds to the typical resolution of the THINGS and HERACLES surveys of nearby disc galaxies \citep{Walter2008, Leroy2009} and is within the range of large integral field spectroscopic surveys (e.g. SAMI \citealt{Croom2012}, CALIFA \citealt{Sanchez2012}, MaNGA \citealt{Bundy2015}), and (\emph{ii}) it fulfills the requirement of the formalism for the $Q_i$ parameters that the scales considered (here, $700 \pc$) are significantly greater than the scale-height of the discs ($100\mh 350 \pc$).

\fig{profilelowres} shows the radial profiles of surface density, velocity dispersion and $Q_i$ for the three components: atomic gas, molecular gas and stars. All the quantities are derived using robust statistics\footnote{Robust statistics are used to provide accurate estimates of statistical quantities for sparse datasets. In this paper, we use the median and the robust standard deviation which is computed as the median absolute deviation (MAD) divided by 0.6745. See \citet{Muller2000} and \citet{Romeo2016} for details.}, they are measured at the scale of $700\pc$, and their profiles are binned at a scale of $1 \kpc$. To limit the numerical noise introduced when computing derivatives, the velocity curve of the galaxy is smoothed before computing $\kappa$ (using a Savitzky-Golay algorithm with a kernel size of 9 data points). 

The density profiles of all components in all the discs are compatible with an exponential form. However, we note that such a profile is imposed in the initial conditions, and therefore the convergence toward the exponential shape cannot be rigorously demonstrated here (but see \citealt{Elmegreen2013, Struck2019}). Despite having a massive central clump, the gaseous components of the \gm and \gl cases yield shallower density profiles than at low gas fraction. The atomic component has a rather uniform distribution, and the radial variations are mostly seen in the molecular phase. The presence of the massive clumps introduces important local variations in the density profiles of \htwo at high \fgas, also seen in the stellar component but with lower amplitudes. Together with \fig[s]{maps} and \ref{fig:psd}, this confirms that the different morphologies caused by varying the gas fraction are mainly affecting the molecular gas and thus the young stars formed in these clumps.

The variations of $\sigma$ with galacto-centric radius contradict the assumption of uniform velocity dispersions in disc galaxies (e.g. \citealt{Obreschkow2016}), and illustrates a process of disc heating \citep{Goldbaum2015, Zhang2016}. This is particularly visible in the gas-rich cases, likely because of radial inflows of matter amplified by the presence of the massive clumps \citep{Noguchi1999, Dekel2013}. Interactions between gas flows and the clumps induce exchanges of angular momentum favoring inflow toward the galactic center. Conversely, in the \gs galaxy with no clumps, a strong bar could in principle play a similar role \citep{Romeo2015, Emsellem2015}, but its absence in our simulation limits the radial inflow in \gs. The profiles of velocity dispersion are comparable to those observed in THINGS, HERACLES and CALIFA galaxies (in \hi, see \citealt{Tamburro2009}, \hi and \htwo, see \citealt{Romeo2017}, and in the stars \citealt{Mogotsi2019}), with values of the other of a few $10 \kms$, and variations of factors $\approx 2\mh 5$ between the inner and outer galaxy. As noted for other quantities, the two gas rich cases have similar behaviors, while the galaxy with a lower gas fraction stands apart.

The values and overall radial trends of the individual $Q_i$ are in qualitative agreement with those derived from observations at comparable resolution \citep[his figure 1]{Romeo2020}, despite being slightly lower (because of surface densities higher than the observed average values). The profiles of $Q_i$ yield similar shapes and values in all our galaxies, showing that the differences in velocity dispersion discussed above are balanced by the gravitational term (surface density). For the gaseous components, the 1-$\sigma$ variations in the profiles are similar in all three galaxies, but $Q_\star$ has a significantly smaller dispersion in \gs than in the gas-rich cases. This is driven by the higher SFR of the latter, where more young stars inherit the dynamical properties of the molecular gas, and increase the contrast with older populations. This however does not influence the overall profiles of $Q_\star$, but only local values, at the positions of the star forming clumps and young star clusters.

By ignoring the relative contribution of the three components to disc instability, this formalism cannot explain why our three cases yield different morphologies and structures, all other things being equal. Although appealing by their simplicity, the single-component stability criteria do not capture the mutual effects of multiple components. While the process favoring instability (gravitation) has an additive behavior (i.e. the masses of the components can be added), the stabilizing agents (rotation and pressure) do not. For instance, a stellar disc and a gas disc could be individually stable ($Q_i \gtrsim 1$), where their combined effect could lead to instabilities \citep[e.g.][]{Lin1966, Jog1984}. This problem has already been illustrated by \citet[][his figure 1]{Romeo2020} using observational data of nearby disc galaxies. Along similar lines, \citet{Westfall2014} reported no relation between the individual $Q_i$ and the star formation activity, indicating that the formation of star forming structures cannot be described without accounting for the interplay of all the disc components (see also \citealt{Leroy2008}). This calls for a multi-component stability approach, as discussed in the next section.

\begin{figure}
\centering
\includegraphics{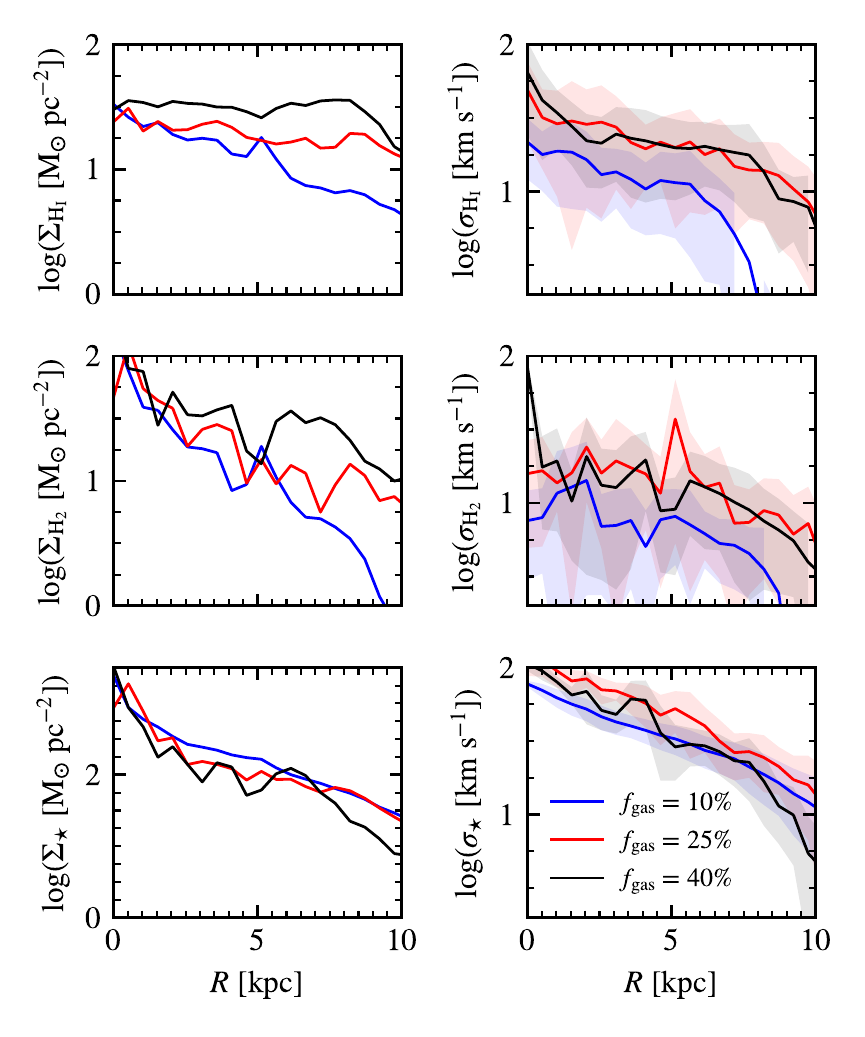}
\caption{Same as \fig{profilelowres}, but with all quantities measured on a scale for $70\pc$, and binned in radial bins of $0.5 \kpc$. The two gas rich cases show local variations due to the massive clumps, less severely averaged out than at low resolution. The Safronov/Toomre $Q_i$ stability parameters are unphysical at this resolution, and are thus not shown here (see text).}
\label{fig:profile}
\end{figure}

More recent observational surveys of the gas structures in nearby disc galaxies are pushing the resolution to several 10s of pc (e.g. PAWS, \citealt{Schinnerer2013}, PHANGS-MUSE, Emsellem et al. \inprep), which calls for a finer analysis of disc stability. To this end, we adopt now the scale of $70 \pc$ to measure the dynamical quantities.

\fig{profile} repeats the measurement of the radial profiles, but now at a resolution of $70 \pc$, and using radial bins of $0.5 \kpc$. As expected the changes in the density profiles are mild, and the conclusions from the analysis at low resolution still hold. However, more significant differences appears in the velocity dispersions. Contrary to the case examined at low resolution (\fig{profilelowres}), velocity dispersions are now lower in the molecular phase than in the atomic gas. However, as seen for the surface density, local variations of $\sigma_{\htwo}$ are caused by the presence of the massive clumps in \gm and \gl. Here again, these variations tend to disappear in the stellar component, where the smooth distribution of the old stars (i.e. well decoupled from their gas nurseries) dominates.

The significantly lower velocity dispersion at high resolution, notably in \htwo, illustrates the problem of beam smearing, i.e. the contamination of the velocity dispersion by the radial dependence of the rotation curve of disc galaxies, which artificially increases the dispersions at low resolution (see \citealt{Bacchini2019, Bacchini2020}, Ejdetj\"arn et al. \inprep).

The new scale of our analysis ($70 \pc$) is well within the disc scale-heights ($\sim 100\mh 200 \pc$), and thus out of the application range of the Safronov/Toomre formalism: it violates the assumption of infinitesimally thin discs. In addition to neglecting the multi-component interplay as discussed above, applying \eqn{qt} at this resolution would lead to unphysical results. Therefore, one must use another, more advanced, stability parameter.

\subsection{Multi-component stability analysis}

For the rest of the paper, we adopt a different formalism for the stability analysis, as motivated by: (\emph{i}) the need to account for multiple disc components, and (\emph{ii}) the requirement of a parameter which is physically valid at scales below the disc scale-heights.

\subsubsection{Stability parameter and characteristic instability scale}
\label{sec:qrf}

To account for the different contributions of multiple components to the net stability regime of a galactic disc, \citet{Romeo2013} introduced the multi-component parameter\footnote{This rigorously derived parameter differs from the \citet{Wang1994} parameter ($Q_{\rm WS}^{-1} = Q_\star^{-1} + Q_{\rm gas}^{-1}$) by accounting for the thickness of the disc (through the term $T_i$, \eqn{t}), and the different weights of the components (through the term $W_i$, \eqn{w}), such that the net local dynamical regime is dominated by the most unstable component. See \citet{Jog1996} and \citet{Romeo2011} for discussions on the inaccuracy of the Wang \& Silk parameter.}:
\begin{equation}
\label{eqn:qrf}
\qrf = \left( \sum_i \frac{W_i}{T_i Q_i}\right)^{-1},
\end{equation}
where $Q_i$ is the stability parameter of the $i$-th component (\eqn{qt}). The term $T_i$ quantifies the stabilization effect due to the thickness of the disc, i.e. it increases the effective stability parameter $T_iQ_i$ of the $i$-th component when the matter is not vertically concentrated \citep{Romeo1994}. It reads
\begin{equation}
\label{eqn:t}
T_i = \left\{ 
\begin{array}{ll}
1+0.6 \left(\frac{\sigma_z}{\sigma_R}\right)_i^2 & \textrm{if} \quad \left(\frac{\sigma_z}{\sigma_R}\right)_i \leq 0.5\\
\\
0.8+0.7 \left(\frac{\sigma_z}{\sigma_R}\right)_i & \textrm{else},
\end{array}
\right.
\end{equation}
(see \citealt{Romeo2013} for details and discussion on the accuracy of this parameter). The expression for $T_i$ accounts for the full diversity of cases observed in stellar discs (see \citealt{Gerssen2012}, \citealt{Pinna2018}, and \citealt{Walo2021} on debated relations between $\sigma_z/\sigma_R$ and the morphological type of the galaxy). In particular, the second case in \eqn{t} depicts systems tending toward highly isotropic kinematics, for instance when supported by dynamical heating and turbulence in the gaseous component and the young stars\footnote{In our galaxies, the ratio $\sigma_z/\sigma_R$ is almost everywhere greater than 0.5, and often greater than unity. This contrasts with observational data of nearby discs, and is probably caused by the high SFRs in our models, and also possibly by numerical heating.}. 

Finally, $W_i$ is used to attribute different weights to the components:
\begin{equation}
\label{eqn:w}
W_i = \frac{2\sigma_m\sigma_i}{\sigma_m^2+\sigma_i^2},
\end{equation}
where $m$ identifies the component which locally dominates the net instability regime, i.e.
\begin{equation}
\label{eqn:m}
T_mQ_m = \min_{i}\left\{T_iQ_i\right\}.
\end{equation}
This formalism gives less weight to the components whose velocity dispersion is very different from the most unstable one. In this paper, we consider the three components $i \in$ \{ \hi, \htwo, $\star$ \} representing the atomic gas, the molecular gas and the stars. 

For both the Safronov/Toomre and the Romeo \& Falstad parameters, a value greater than unity denotes stability, but only against axisymmetric perturbations, which explains why some galaxies are observed to have $Q \gg 1$ \citep{Romeo2017}. Invoking a higher threshold ($\gtrsim 2\mh 3$) is necessary to account for non-axisymmetric perturbations \citep{Griv2012}, as well as for the dissipative nature of the gaseous components \citep{Elmegreen2011}.

Once the component locally driving the instabilities is identified (as $m$, using \eqn{m}), we compute the characteristic instability scale, i.e. the size of the region in which gravitational disc instabilities develop, as
\begin{equation}
\label{eqn:lrf}
\lrf = \frac{2\pi \sigma_m}{\kappa}.
\end{equation}

Contrary to the Safronov/Toomre $Q_i$, the \qrf parameter retains a physical meaning at scales below the disc scale-height, since it is based on a dispersion relation that converges at small scales towards the Jeans regime of instability including rotation. Yet, it relies on the short wavelength and the epicyclic approximations \citep{Binney2008, Romeo2013}. \app{approx} shows that our simulations fulfill these important requirements in all regions of the galaxies, which validates these approximations.

\begin{figure}
\centering
\includegraphics{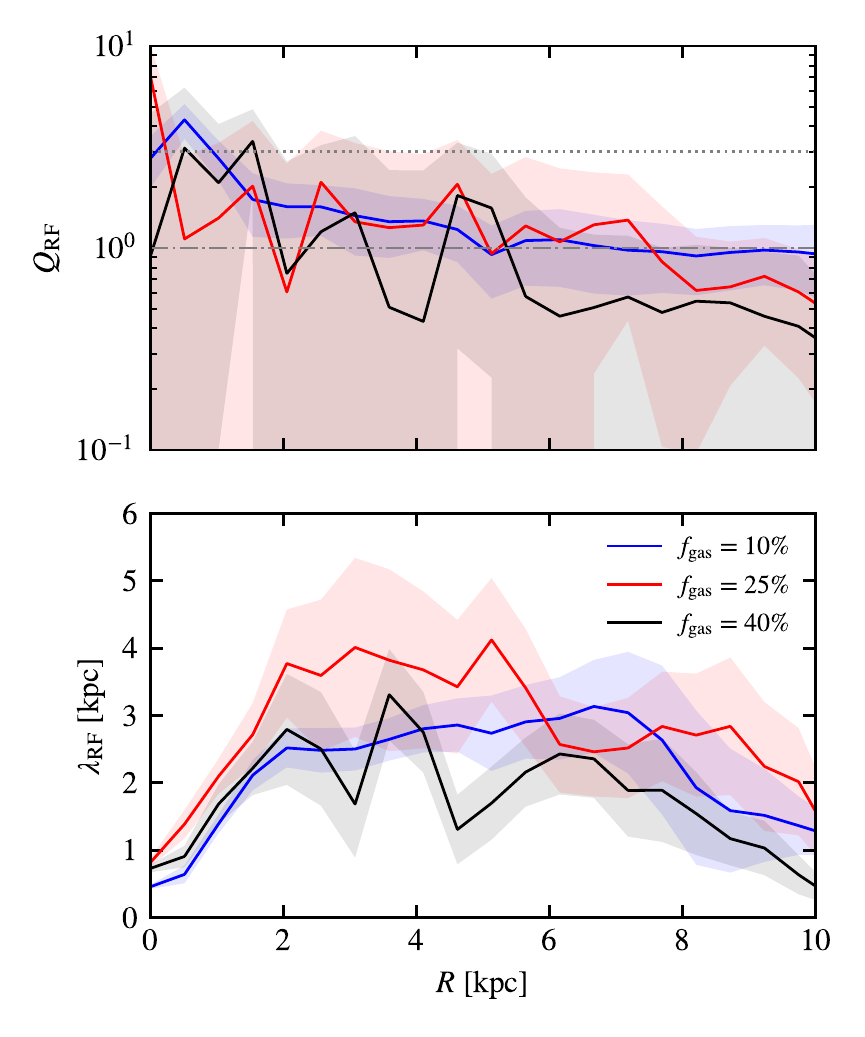}
\caption{Radial profiles of the \qrf and \lrf parameters (\eqn[s]{qrf} and \ref{eqn:lrf}), using the same resolution and binning as in \fig{profile}, i.e. $70 \pc$ and $0.5 \kpc$. The horizontal lines indicate the approximate boundaries for stability against axisymmetric ($\lesssim 1$) and non-axisymmetric perturbations ($\lesssim 3$).}
\label{fig:qprofile}
\end{figure}

\fig{qprofile} shows the radial profiles of \qrf and \lrf. \qrf shows that the discs are marginally stable, with almost no radial dependence. We note an overall decrease of \qrf with increasing \fgas at all radii. Combined with the information on the surface density and velocity dispersion from \fig{profile}, this confirms the expectation that the gaseous component plays an increasing role on the stability regime at high \fgas. However, the three galaxies yield \qrf's of the same order of magnitude and within each-other's dispersions, which mitigates this importance of \fgas. The explanation for such an unambiguous, yet mild, influence of the gaseous component on the stability regime is given in \sect{drivers}.

\qrf of the gas-rich cases show local, stochastic variations caused by the massive clumps, while that in \gs is smoother, and with a significantly smaller 1-$\sigma$ dispersion. The dispersion in the former represents the high degree of asymmetry in the morphologies of the gas-rich discs, where massive clumps and large under-dense regions can be found at the same radii, and are thus blended in the same radial bin in \fig{qprofile}. Although using the multi-component parameter $\qrf$ significantly improves the (at least qualitative) agreement between the stability analysis and the morphologies, the large scatter shows the limitations of our one-dimensional diagnostic, and calls for a fully local two-dimensional analysis, as shown below.

In terms of \lrf, the three cases yield instability scales of $\approx 2\mh 4 \kpc$ over a large range of radii. Apart from stronger variations and dispersions in the gas-rich cases, the overall profile of \lrf does not significantly change, and in a non-systematic way, with \fgas. As noted above, the size of clouds in the \gs case is comparable to the diameter of the spiral arms and the scale-height of the disc ($\approx 100 \pc$). However, the sizes of some clumps in \gm and \gl are significantly larger the disc scale-height. This demonstrates that the size of the structures detected in discs (clouds or clumps) are not exclusively set by large-scales instabilities, and that additional processes must be invoked, likely during the collapse and assembly phase of the clouds. These processes are highly non-linear, and thus cannot be included in a formalism like the one adopted here. Nevertheless, our results suggest different regimes of instabilities between low and high gas fractions, at least at the scales of their main morphological differences, i.e. below a few $100 \pc$.

\begin{figure*}
\centering
\includegraphics{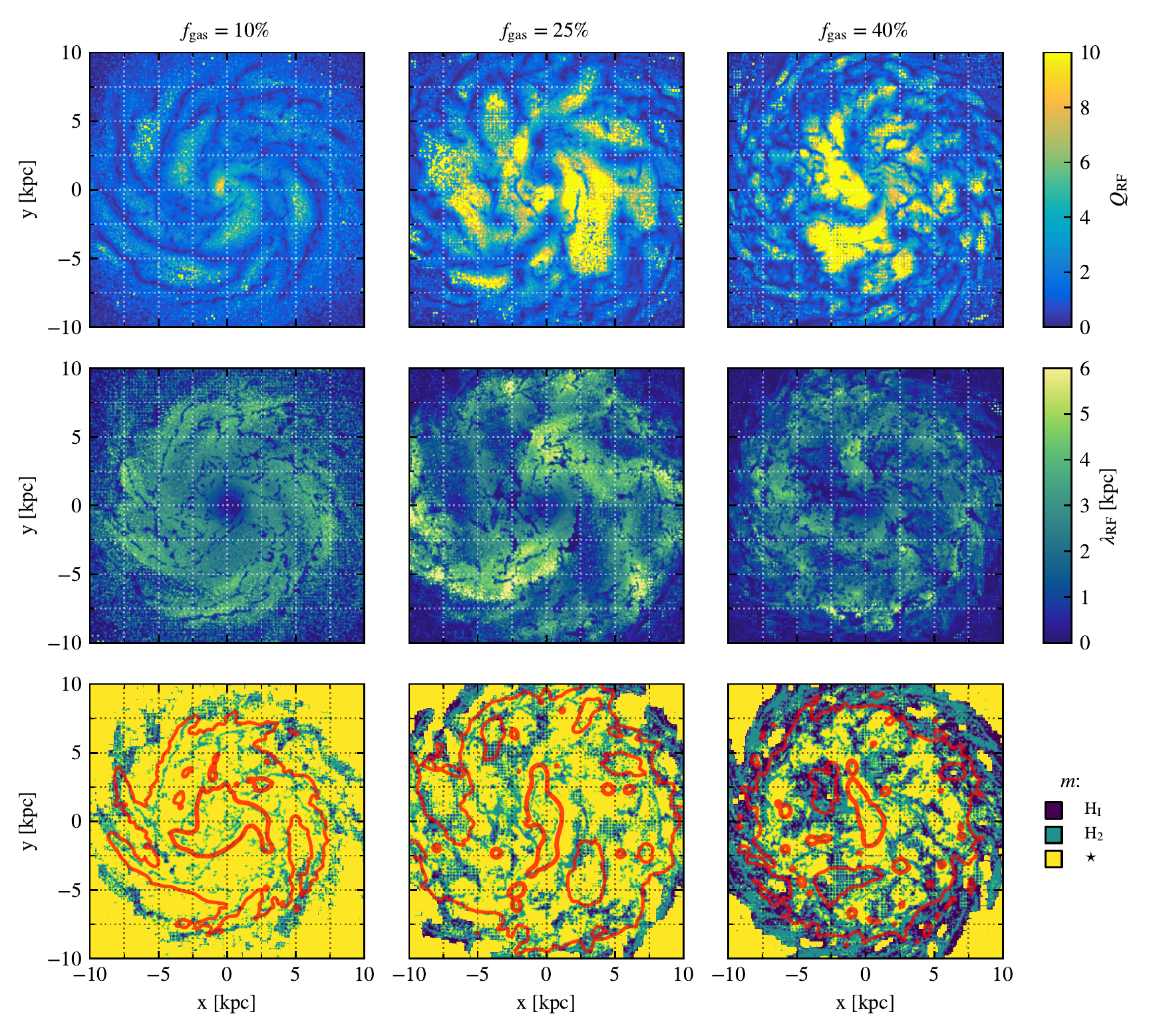}
\caption{Maps of the \qrf stability parameter (\eqn{qrf}, top), of the instability scalelength \lrf (\eqn{lrf}, middle), and of the component dominating the instability regime (identified by the parameter $m$, \eqn{m}, bottom). All quantities are computed at a scale of $70\pc$. Contours in red indicate stellar surface densities of 3 and $300 \Msun\ \pc^{-2}$, to guide the eye.}
\label{fig:qmaps}
\end{figure*}

\begin{figure}
\centering
\includegraphics{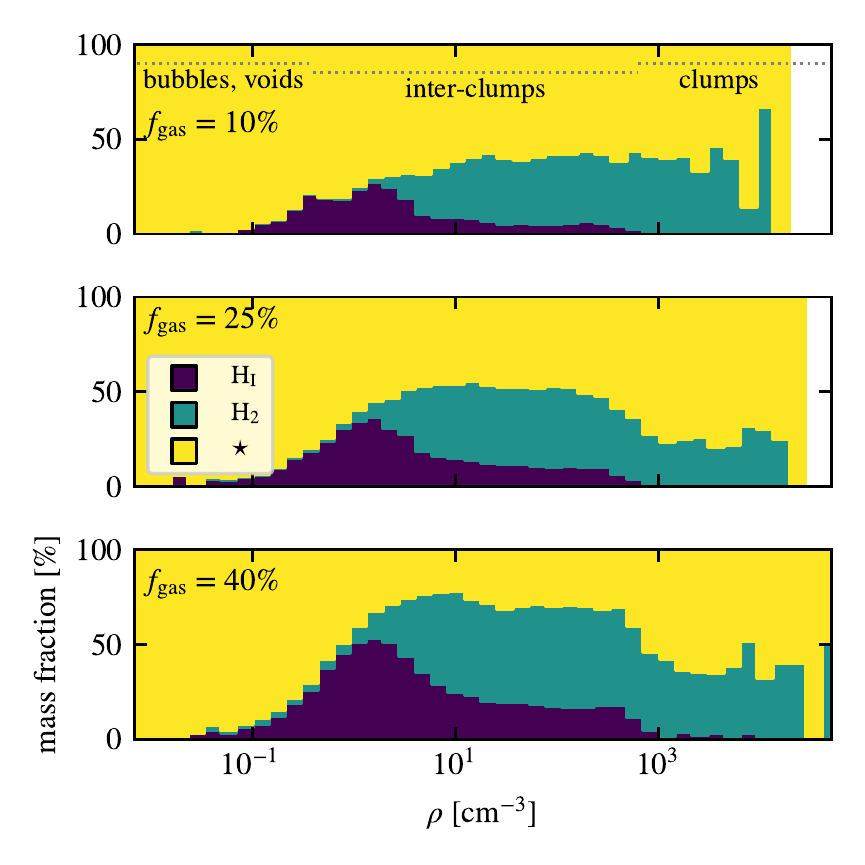}
\caption{Fraction of the gas mass in regions where the instability is set by the atomic, the molecular and the stellar component, as a function of the gas density. Approximative distinctions between different media are shown at the top, for reference. While most of the mass is in stellar-driven instability regions, this fraction strongly depends on the density, and the gas fraction of the galaxy.}
\label{fig:mpdf}
\end{figure}

As noted by \citet{Romeo2017} with observational data, the radial variations of $\sigma_i$ (\fig{profile}) have a negligible impact on \qrf, which remains flat over a large radial range. These authors however noted an important effect on the size of the unstable regions. This further calls for a local study to complement our analysis using 1D radial profiles. To this end, the top and middle rows of \fig{qmaps} shows the maps of \qrf and \lrf computed at a scale of $70 \pc$. These maps illustrate the azimuthal variations lost in the one-dimensional radial profiles. Within the discs ($\lesssim 10 \kpc$), the lowest values of \qrf coincide with the structures in molecular gas. In \gs, this corresponds to the spiral arms and the clouds they host, as expected. In this galaxy, this is also where \lrf reaches its local minima, with instability scalelengths shorter than 1 \kpc. The structure of this unstable medium highlights a characteristic size of $240 \pc$ (as found on a PSD of the maps of \fig{qmaps}, not shown), set by the separation of the spiral arms.

The \gm and \gl cases show a number of features at scales greater than $1 \kpc$, notably at high \qrf, and long \lrf. These correspond to areas of low density between the clumps. Such under-densities result from disc dynamics (e.g the inter-arm medium) and the vigorous stellar feedback that affects the distribution of gas at these scales. While the former is set by gravity and thus affects both the stars and the gas, the latter carves bubbles in the gas only, leaving the stellar component mostly unaffected. As a consequence, these two effects leave different signatures on \qrf and \lrf: in under-densities caused by large-scale dynamics, both the stellar and gas densities are low, as well as the radial velocity dispersions. As a result, \qrf tends to be high while \lrf remains short (e.g. at $x,y \approx -3\kpc, 3\kpc$ in \gm, and $-2 \kpc, 2 \kpc$ in \gl). This corresponds to long timescales of the secular, adiabatic evolution of the disc ($\sim 100\mh 1000 \Myr$). Conversely, feedback bubbles are regions of low density gas, but without any particular deficit of stars. There, the expansion of the bubble and the turbulence injected by feedback increase the velocity dispersion isotropically, and thus also in the radial direction. Combined with the reduced density of gas, and the moderately-affected density of stars, this leads to both \qrf and \lrf being high (e.g. at $x,y \approx 2\kpc, 6\kpc$ in \gm, and $0\kpc, -2\kpc$ in \gl). The timescales associated with stellar feedback are usually significantly shorter than those associated with disc dynamics, and such differences could thus translate into different instability regimes, as illustrated in \sect{decoupling}.

By using the two-component (gas and stars) framework for stability analysis from \citet{Romeo2011}, \citet{Inoue2016} noticed that the material assembling into clumps lies in regions with $Q > 1$. This results is not directly comparable to ours, as it only concerns the clump material traced back to its origin before the formation of the clump, while we characterize the instabilities over the entire disc, independently of the fate of the gas. The gas that assembles into clumps evolves through a range of stability states, as gas recycling is set by disc dynamics and feedback. The timescales of these cycles vary with local and global conditions (density gradients, coupling scale of feedback to the ISM, shear etc., see \citealt{Semenov2017}), such that the moment of the onset of clump formation varies from clump to clump, and from galaxy to galaxy. The (necessary) simplification of measuring the $Q$ parameter at an arbitrary epoch before the assembly of the clumps also supposes that clumps form as the monolithic collapse of a single gas region, which might not be correct for all clumps (\citealt{Behrendt2019}, Renaud et al. \inprep).

By not separating the atomic and molecular phases, the $Q$ parameter used in \citet{Inoue2016} under-estimates the contribution of the molecular gas to the stability (see \citealt{Romeo2013}). Although this is probably negligible in the inter-arm medium at low gas fractions (which is almost exclusively atomic), it is not the case in the medium hosting the formation of clumps, i.e. the spiral arms at low \fgas, and the inter-clump medium of gas-rich discs (see the next section). Our results confirm that material in between clumps can be found with $\qrf < 1$, thus very likely the seeds of future clumps.

To explain their finding of $Q > 1$ in future clump material, \citet{Inoue2016} invoke external processes, of cosmological origin, like minor mergers, stream accretion and tidal compression as non-linear processes missing in the theoretical formalism, to explain the discrepancy with their simulations. While such effects are present in galaxies, particularly at high redshift, our study using isolated galaxies shows that they are not necessary to explain the formation of massive clumps, and that internal instabilities are sufficient to drive the formation of massive clumps.

\subsubsection{Drivers of instabilities}
\label{sec:drivers}


The last row of \fig{qmaps} shows the maps of $m$, identifying the component which dominates the instability regime (\eqn{m}). \fig{mpdf} complements this analysis by showing, as a function of gas density, what fraction of mass lies in regions where the stability regime is driven by each component. At least $\approx 50\%$ of the gas mass is in regions of $m = \star$, at virtually all densities in the \gs and \gm galaxies. This confirms the prominent role of stars in driving instabilities in nearby spiral galaxies \citep{Romeo2016, Romeo2017, Marchuk2018}. The next section shows that the majority (but not all) of these stellar-driven instabilities are coupled with the gaseous phase, and thus that the stars can drive gaseous instabilities.\footnote{The formalisms of the stability parameters do not capture the full range of perturbations (e.g. non-axisymmetric), and thus the actual value of \qrf cannot be quantitatively linked to the collapse of structures nor to the star formation activity. An approximate threshold of $\qrf \sim 2\mh 3$ is used to distinguish between stable ($\qrf \gtrsim 2\mh 3$) and unstable ($\qrf \lesssim 2\mh 3$) regions. The precise value of the critical stability level is still questioned \citep{Romeo2015}, since it is influenced by complex phenomena such as non-axisymmetric perturbations and gas dissipation, whose effects are difficult to quantify (recall \sect{qrf}). We rather focus here on relative differences between galaxies and media, in particular on which component is the least stable one.}

Not surprisingly, the mass fraction in $m=\star$ decreases with increasing \fgas, down to $\approx 25\%$ in the inter-clump medium of \gl ($\sim 1\mh 1000 \cc$), where gas instabilities take over, especially in the molecular phase. The inter-clump medium containing, by definition, the formation sites of future clumps, the dependence of the driver of instabilities on \fgas reflects the diversity in the nature and scale of clumps between our three galaxies.

Perhaps more surprisingly, in the densest medium, corresponding to the clouds/clumps themselves ($\gtrsim 10^3 \cc$), the molecular phase drives instabilities for $40 \%$ of the gas mass in \gs, and only $20\mh 25\%$ in \gm and \gl, the rest being driven by stars (due to the absence of atomic gas at such high densities). These low fractions are caused by the high level of turbulence stabilizing the inner regions of the gas clumps, especially in the gas-rich cases, with Mach numbers $\gtrsim 10$ (at the scale of $70 \pc$). There, the SFR is $1.5\mh 2.0$ times higher in regions where instabilities are driven by stars than by molecular gas, showing that the high levels of turbulences are caused by the feedback from young stars within the massive clumps. Such differences are however not present in \gs, probably because of a weaker SFR, and a different response of the smaller star forming clouds to feedback (Renaud et al. \inprep). Enhanced turbulence leads to $\sigma_{\htwo} > \sigma_{\star}$ and, in turn, to $Q_{\htwo} > Q_{\star}$, without altering $T_i$ due to the isotropic nature of turbulence. In short, feedback from young star forming clumps favor regimes of instabilities driven by the stellar component, while gas-dominated clumps (i.e. at an earlier stage of their evolution, in their assembly phase before they form stars) are in instability regimes driven by the molecular gas.

\begin{figure*}
\centering
\includegraphics{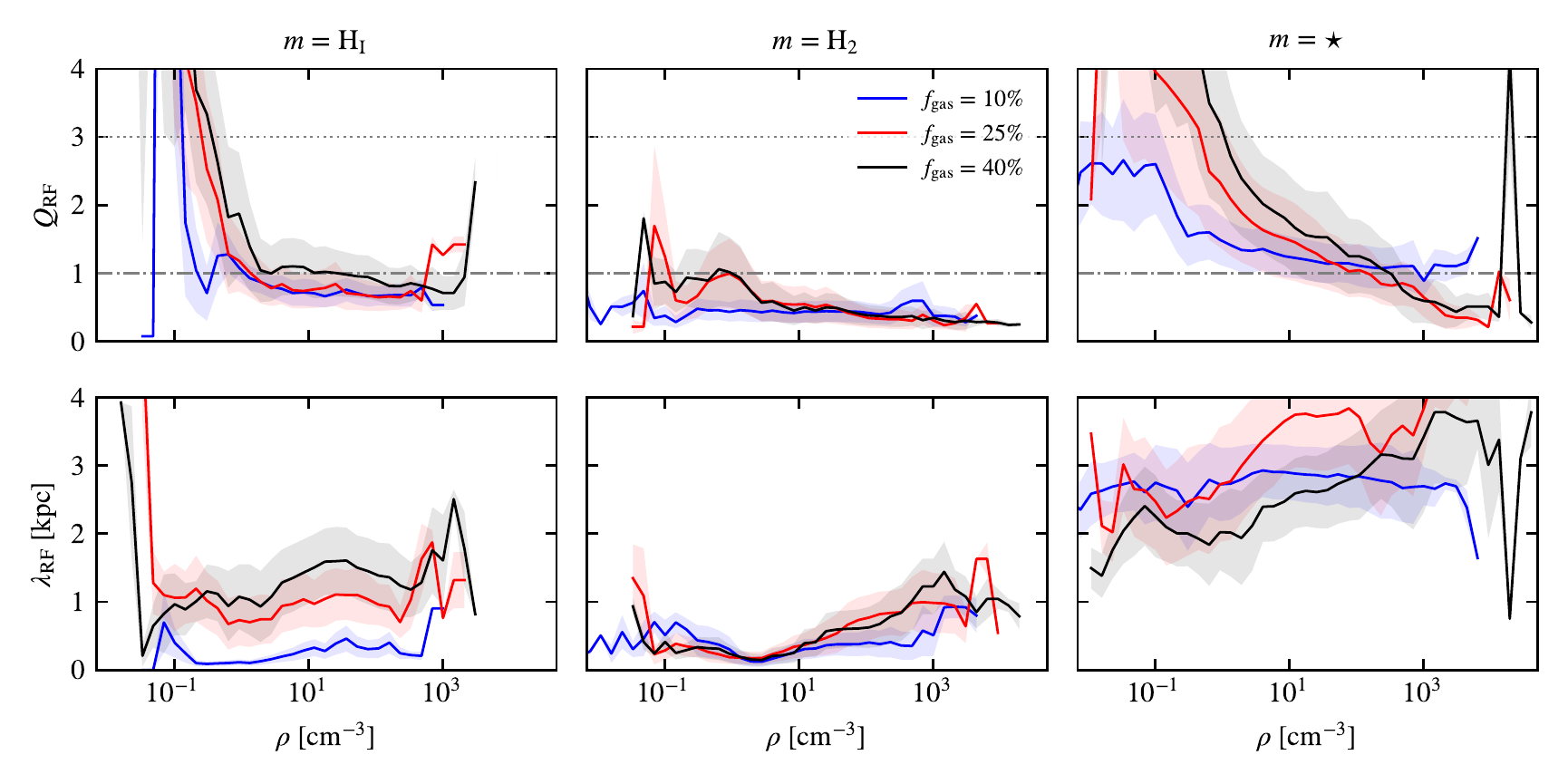}
\caption{As function of the gas density, median \qrf (top) and \lrf (bottom) in regions where the instabilities are driven by the atomic gas (left), the molecular gas (center) and the stars (right). The shaded area indicates the 1-$\sigma$ robust dispersion. Values at the edges of the density range shown suffer from low-numbers statistics. Clear dependences on the gas fraction are mostly visible in the gas-driven instabilities (left and center).}
\label{fig:lpdf}
\end{figure*}

Finally, \fig{lpdf} shows the median \qrf and \lrf as functions of gas density, separating the drivers of instabilities. The dependences on \fgas are mostly visible in the regions with gas-driven instabilities. There, higher gas fractions translate into larger \qrf and \lrf, at virtually all densities, but particularly in the inter-clump medium ($\sim 1 \mh 1000 \cc$). As shown in \fig{mpdf}, gaseous drivers are particularly important in this density range, corresponding to the material potentially available for the assembly of future clumps, and/or to fuel existing clumps with gas. Even in the most gas-rich cases, the values of \qrf at these intermediate densities depict unstable regimes, and thus promote the assembly and collapse of gas structures. Furthermore, the increase of \lrf with \fgas shows that the instabilities appear at larger scales in gas-rich cases, suggesting that the assembly of massive clumps is favored at high \fgas (as opposed to smaller clouds at low \fgas). However, we recall that \lrf is not a direct indicator of the size of future clumps, and that other, non-linear, regulation mechanisms intervene during the later stages of their formation. While the size of the clumps remains difficult to predict, the existence of large \lrf at high \fgas, yet with similar density PDFs between the three galaxies in this regime (\fig{pdf}), indicates that a larger \lrf translates into more massive clumps. Here again, the similarities in \lrf and the density PDF in the inter-clump medium of the two gas-rich galaxies (\gm and \gl) explains the resemblances in the masses of their clumps (see Renaud et al. \inprep).

In the regions with $m=\star$, the picture is more complicated. At low densities (where stars are the main drivers of instability, recall \fig{mpdf}), high values of \qrf denote stable (in \gm and \gl) or marginally stable (\gs) media. In the inter-clump density range ($\sim 1 \mh 1000 \cc$), stellar-driven instabilities concern a smaller fraction of the mass, but lower values of \qrf account for the unambiguous instabilities with $\qrf < 2\mh 3$ (i.e. below the stability threshold, as noted above), as also seen in the clumps themselves ($\gtrsim 10^3 \cc$). However, the associated scalelengths show no clear dependence on the gas fraction, further highlighting the role of the gas phase in driving the differences between molecular clouds and massive clumps.

To summarize, the stellar-driven instabilities concern the majority of the surface (\fig{qmaps}) and the mass (\fig{mpdf}) in all galaxies. The weakening, yet persistence, of this dominance even at high \fgas explains the clear, yet mild, global trend of \qrf decreasing with \fgas noted in \fig{qprofile}. However, it is preferentially the gas-driven instabilities that lead to the formation of the star forming clumps from the gas at intermediate densities, and the scale of such instabilities correlates with the gas fraction.

\subsubsection{Decoupling}
\label{sec:decoupling}

\begin{figure}
\centering
\includegraphics{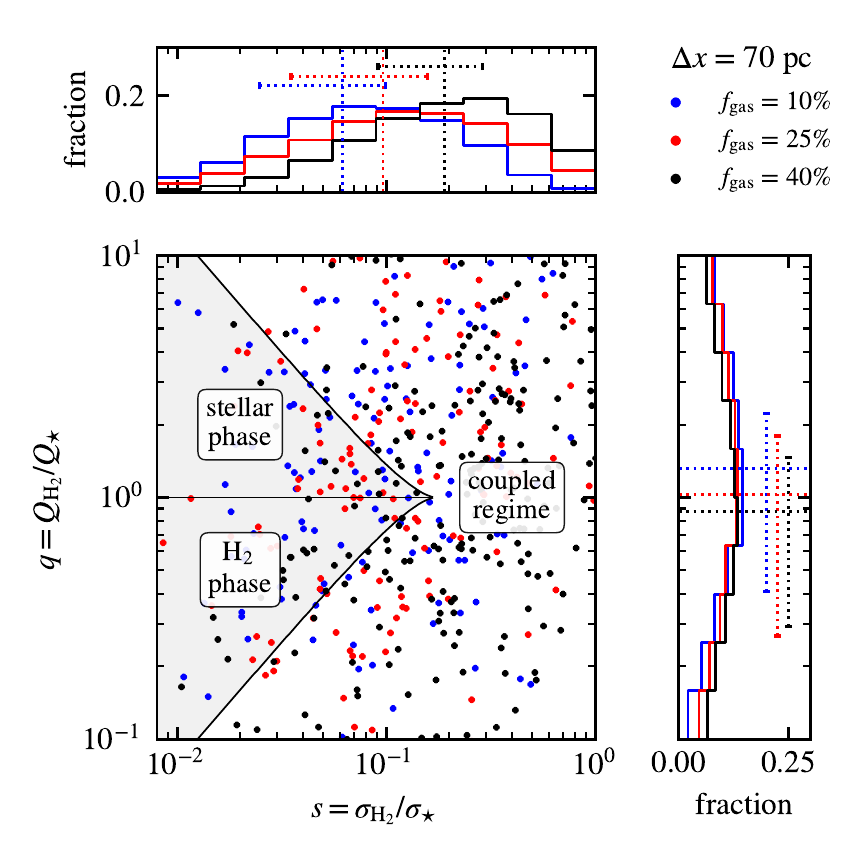}
\caption{Parameter plane of gas and star dominated instabilities, with quantities measured in $70 \pc \times 70 \pc$ beams. For clarity, only 1\% of the points are shown. The shaded area at small $s$ indicates the regime where the stellar and the molecular component are decoupled (see \citealt{Romeo2013} for details). In the other regime, a given component responds to the instabilities driven by the other. Histograms in the top and right-hand panels indicate the distribution of regions for our three gas fractions, with dotted lines marking the median values, and errors bars showing the robust standard deviation.}
\label{fig:decoupling}
\end{figure}

\begin{figure*}
\centering
\includegraphics{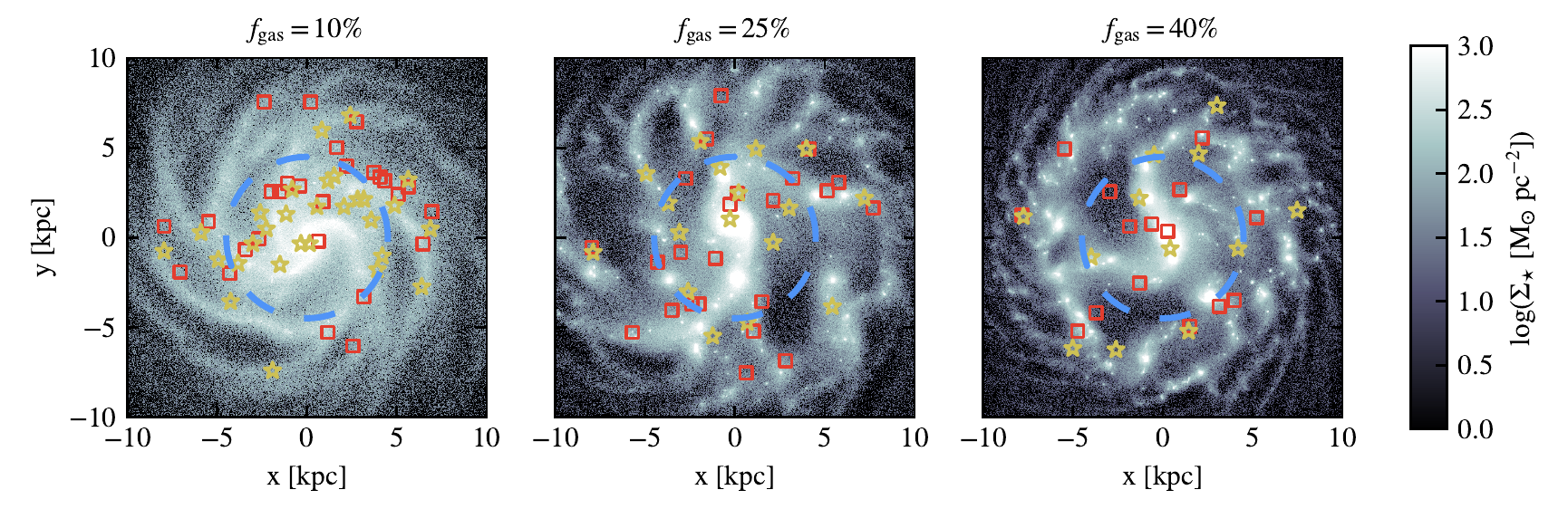}
\caption{Location of the $70\pc \times 70\pc$ regions where instabilities are decoupled, and driven by the molecular gas (squares) and stellar phases (stars). Only 1\% of the points are shown, for clarity. The cyan circles show the radius selected for the analysis of \sect{ab}.}
\label{fig:maps_decoupled}
\end{figure*}

From the analysis above, it remains to be established if instabilities driven by one component can affect the others. \fig{decoupling} shows the locii of the $70 \pc \times 70 \pc$ regions in the parameter plane for gas and star instabilities, i.e. the ratio of individual $Q$ parameters for the molecular and stellar components, versus that of their velocity dispersions. In the shaded area, a component is dynamically too hot to respond to the instability driven by the other and is thus decoupled from the one driving the instability (\citealt{Bertin1988, Romeo2011}, see \citealt{Romeo2013} for details). 30\%, 18\% and 9\% of the regions lie in the decoupled regime for \fgas = 10\%, 25\% and 40\% respectively, with approximately half of them in the stellar-dominated phase in each galaxies. 

All the decoupled regions lie in the intermediate range of densities ($0.1 \mh 500 \cc$). By comparing with the galactic-wide means at the same densities, we find that the decoupled regions yield a higher Mach number (by a factor $\approx 1.5\mh 2.0$), in all galaxies. Similarly, the median \qrf and \lrf are slightly higher in the decoupled regions (by factors $\approx 1.2\mh 2.5$ and $\approx 1.3\mh 1.8$ respectively). Following the discussion in \sect{qrf}, this suggests that feedback plays an important role in the decoupling mechanism. Yet, the decoupled regions also yield a higher average shear\footnote{We evaluate the shear as the quadratic sum of the off-diagonal terms of the Jacobian matrix of the velocity field, computed with first-order finite differences over the scalelength of $70 \pc$. This quantity can be interpreted as the squared rotational frequency term of a local centrifugal force. See \citet{Renaud2015d} for details.} than the galactic-wide averages (by a factor $\approx 1.7\mh 2.8$), which implies a contribution from large-scale dynamics too, like the differential rotation of the discs.

To explore this hypothesis further, \fig{maps_decoupled} shows the position of the decoupled regions in the galaxies. In all three cases, they are found at the edge of dense regions (e.g. spiral arms and massive clumps). Furthermore, a visual inspection reveals that a majority of the decoupled regions are preferentially found on the trailing side of these dense structures (all galaxies spin counter-clockwise). The only exceptions are decoupled regions within the central kpc, i.e. in a special dynamical environment not discussed here. In \gs, we visually find more decoupled regions near the portions of spirals with a high pitch angle, but the absence of clear spiral structures in the gas-rich cases forbids firmly drawing comparable conclusions in these galaxies. \citet{Renaud2014} showed that high pitch angles favor the organization of star forming structures in beads-on-a-string along spiral arms, as opposed to spurs or feathers formed by Kelvin-Helmholtz instabilities at low pitch angles. Because of asymmetric drift, young stars from these regions tend to preferentially inject feedback on the trailing sides of the spirals \citep[][their figure 14]{Renaud2013b}. In other words, the role of large-scale dynamics noted above is to favor an efficient injection of feedback in specific regions, but it is ultimately the feedback itself that decouples the gas and stellar components.

However, most of the regions lie in the coupled regime of instability, i.e. with one component experiencing instabilities driven by the other. The projected distributions of points shown in the side panels of \fig{decoupling} indicates a correlation between $s=\sigma_{\rm H_2} / \sigma_{\star}$ and \fgas, compatible with the findings of \citet{Burkert2010} and \citet{Krumholz2010b}. The opposite trend exists with $q = Q_{\rm H_2} / Q_{\star}$, despite larger scatters. This further confirms that, while the importance of the stellar component in driving instabilities anti-correlates with \fgas, the high number of decoupled regions at low \fgas (i.e. a large volume where stars are not directly affected by the gaseous instabilities, and thus by the variations of \fgas) explains the only weak evolution of the overall instability regime with \fgas (as noted in \fig[s]{mpdf} and \ref{fig:qmaps}).

\subsection{Effects of resolution and beam smearing}
\label{sec:resolution}

\begin{figure*}
\centering
\includegraphics{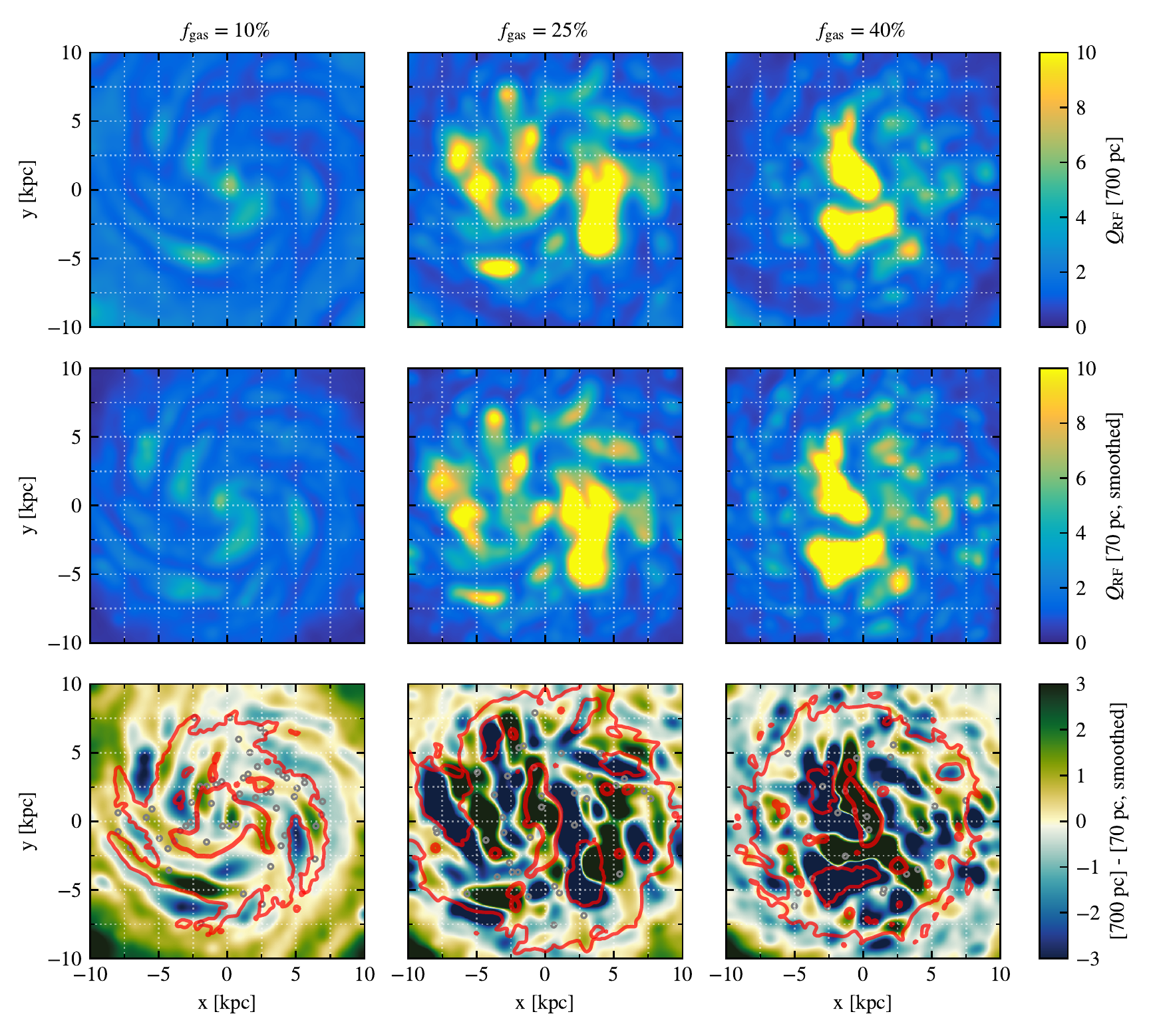}
\caption{Maps of \qrf measured at a scale of 700 pc (top), at a scale of 70 pc and then convolved with a Gaussian with a width of 700 pc (middle), and difference between the two (bottom). Contours in red indicate stellar surface densities of 3 and $300 \Msun\ \pc^{-2}$, to guide the eye. The small gray circles indicate the positions of the $70 \pc \times 70 \pc$ regions where the instabilities are decoupled (as in \fig{maps_decoupled}).}
\label{fig:smooth}
\end{figure*}

As shown in \fig[s]{profilelowres} and \ref{fig:profile}, measuring quantities (densities, velocity dispersions) at large scales implies that distinct physical regimes could be blended together, thus skewing the results. Beam smearing is a well-known example of such a distortion caused by the contamination of the velocity dispersion by the differential rotation of discs (\citealt{Davies2011, Bacchini2019, Bacchini2020}, Ejdetj\"arn et al. \inprep).

To assess the severity of this problem, in this section we repeat our analysis of the instability regime of our models but using the quantities measured at scales of $700 \pc$ (recall \fig{profilelowres}). \fig{smooth} compares the maps of \qrf measured at $700 \pc$ resolution, with those measured at $70 \pc$ and then convolved with a Gaussian kernel of $700 \pc$\footnote{For consistency, the maps at $700 \pc$ are also convolved with the same Gaussian, to account for the non-zero effects of the wings of Gaussian distributions.}. Doing so allows us to eliminate the effects of resolution (beam size), and rather identify the regions where the differences indicate a scale-dependence of the physics of instabilities. The same exercise applied to \lrf is shown in \app{lowres}.

At first sight, the two sets of maps seem comparable as they show features with approximately the same morphologies and sizes. However, the maps of the differences (shown in the bottom row of \fig{smooth}) highlight important local variations. When focusing on the galactic disc (i.e. within the central $\approx 10 \kpc$), the amplitude of the differences and the surface they cover increase with \fgas. In \gs, we find no clear correlation between the amplitude of the difference and the presence of structures like spiral arms and clouds (which host the lowest values of \qrf, as noted in \fig{qmaps}), suggesting big similarities in the physics of instabilities between the 70 and $700 \pc$ scales. In other words, instabilities at low-gas fraction are mainly set by galaxy-scale dynamics (at least down to scales of $70 \pc$). At higher gas fraction however, the maps of \fig{smooth} yield important differences, in particular in the dense clumps, and irrespectively of the underlying values of \qrf. This indicates that different physical regimes are probed when examining the discs at $70$ and $700 \pc$, and thus suggests a transition scale between our two resolutions. We examine and quantify this in \sect{ab}.

In all cases, the regions where the instabilities are decoupled are almost exclusively found in the areas of small differences between the two sets of maps. This confirms that the decoupling is not an artifact caused by our choice of resolution, but instead results from a real physical process. The preferential locations of these regions on the edges of large features seen on the bottom row of \fig{smooth} indicate that the decoupling is closely associated with the transition between the two instability regimes suggested above. In \sect{decoupling}, we showed that feedback plays a more important role in the decoupling than dynamical effects. Therefore, the change of instability regime is likely driven by hydrodynamics.

In summary, our approach shows that the differences between high resolution smoothed by large Gaussian kernel and low resolution are not only differences in the scatter around the mean quantity within each beam, but also a fundamental difference of the mean itself. As a result, smoothing out the fluctuations by increasing the beam size does not lead to the same values as directly measuring quantities over large beams. Therefore, this indicates that differences between high and low resolution are non only caused by small-scale variations within the beams, but also by fundamental physical differences between the instability regimes at small and large scales.

\section{Scale-dependence of the instability regimes}
\label{sec:scale}

\subsection{Methodology}
\label{sec:instamethod}

To evaluate the scale-dependence of the instability regime in the discs, we center our analysis on the galacto-centric radius $R_0 = 4.5 \kpc$, as indicated in \fig{maps_decoupled}. This radius is selected to probe a large variety of structures. To measure the relevant physical quantities, we select the volume between $R_0 - l/2$ and $R_0 + l/2$, and of height above the disc plane between $-l/2$ and $+l/2$, where $l$ is the scale considered in the analysis. In this volume, we define $N = \lfloor 2\pi R_0 / l \rfloor$ cylinders of diameter $l$ and height $l$, with their centers uniformly distributed in azimuth along the galactic circle of radius $R_0$. The cylinders do not necessarily coincide with sub-structures in the discs like clumps or spirals. In each cylinder, we measure the surface density $\Sigma$ and the radial velocity dispersion $\sigma$ of the gaseous and the stellar component. We then compute the median and robust standard deviation of the $N$ values of these quantities. Note that if a majority of the cylinders do not contain mass, as it is the case at small scales in \htwo, these medians are null. We repeat the process by varying the scale $l$ between $l_{\rm min} = 40 \pc$ and $l_{\rm max} = 4 \kpc$. Finally, we compute the derivative of the relations of the medians of $\Sigma$ and $\sigma$ with the scale $l$ to obtain the indexes $a$ and $b$ defined as 
\begin{equation}
\label{eqn:ab}
\Sigma \propto l^a \qquad \textrm{and} \qquad  \sigma \propto l^b.
\end{equation}

This approach is similar to that in \citet{Agertz2015b} who applied it to galaxies with low gas fractions. To avoid biases in our measurements, the cylinders are not centered on specific structures like clumps. In fact, since the inter-clump medium represents a large fraction of the surface of the disc, the cylinders are more likely found in between dense structures. Then, we expect high values of $a$ at scales where cylinders start to encompass clumps, especially in the molecular phase which shows stronger density contrasts than the smoother distributions of atomic gas and stars (recall \fig{psd}).

\subsection{Instability regimes}
\label{sec:ab}

\begin{figure*}
\centering
\includegraphics{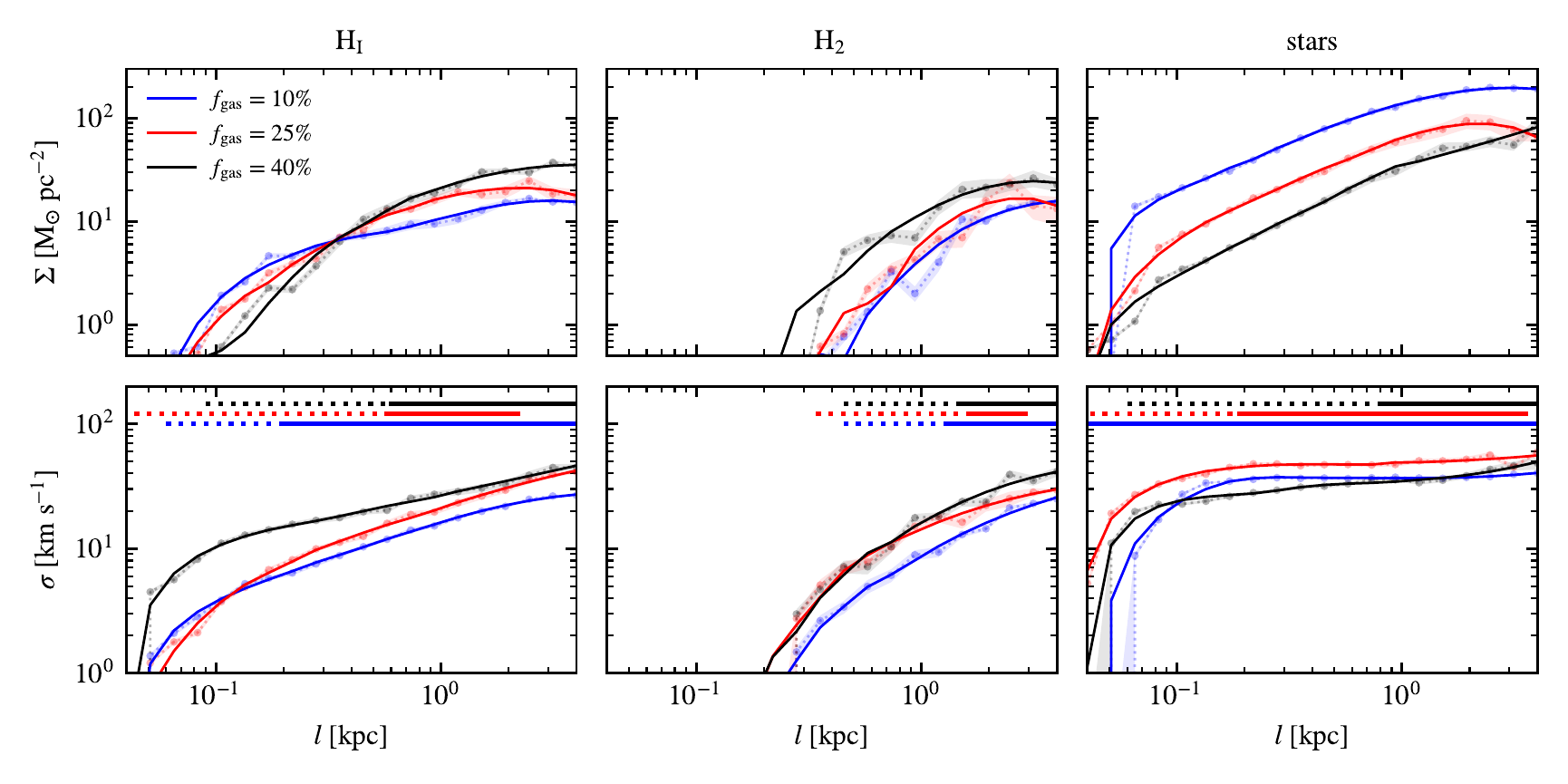}
\caption{Surface density (top) and radial velocity dispersion (bottom) of the gas (left) and the stars (right), as functions of scale $l$. Raw data is shown by points, with the shaded areas indicating the uncertainties, accounting for the sample size (i.e. the number of cylinders $N$). Solid lines are a smoothed version, for clarity. The solid horizontal lines indicate the range of scales under regime A of instability (Toomre-like), and the dotted line show that under the regime D. In all the components of all the galaxies (except for the stars in \gs), a transition between a disc-driven and a clump-driven regime of instability occurs at scales of a few $100 \pc$.}
\label{fig:insta}
\end{figure*}

Based on the values of $a$ and $b$, \citet[see also \citealt{Romeo2014, Agertz2015b}]{Romeo2010} identified a set of instability regimes, the most important of which are briefly presented here.
\begin{itemize}
\item Regime A: $a < 1$ and $b < (a+1)/2$. Pressure drives the evolution of the disc at small scales, gravitation at intermediate scales, and rotation at large scales. This corresponds to Toomre-like instabilities.
\item Regime A': $a > 1$ and $b > (a+1)/2$. Similar as before, as Toomre-like instabilities, but with the scales of dominance of pressure and rotation reversed.
\item Regime B: $b = (a+1)/2$. The gravitation and pressure terms have the same dependence on scale, and thus the medium follows the virial scaling (i.e. $\sigma^2 \propto l \Sigma$).
\item Regime C: $a < 1$ and $b > (a+1)/2$. Self-gravity dominates the instability towards small scales, while pressure and/or rotation tend to stabilize the large scales (e.g. a cold, non-turbulent disc).
\item Regime D: $a > 1$ and $b < (a+1)/2$. Self-gravity dominates the instability towards large scales, while pressure and/or rotation tend to stabilize the small scales (e.g. a non-rotating, non-turbulent sheet). 
\end{itemize}

\fig{insta} shows the dependence of the surface density and radial velocity dispersion with scale, and identifies the scales under regime A (Toomre-like). As reported by previous works \citep{Romeo2010, Romeo2014, Agertz2015b} and suggested by many of the results presented above, the instability regime evolves between the small and large scales. In our simulations, the details of this evolution are complex and vary between galaxies and components. However, we can highlight the general trend that the components lie in regime A at large scales, and transition to regime D at smaller scales. The transition scale between the two regimes varies with \fgas for the atomic and stellar components. At low gas fraction, these components are in the Toomre-like regime down to small scales ($\sim 200 \pc$ in \hi, and down to our resolution limit in stars). At larger \fgas, this transition occurs at larger scales ($\approx 600 \pc$ in \hi, and $\approx 200 \pc$ and $\approx 800 \pc$ in stars for \gm and \gl respectively). This demonstrates that the transition between disc instabilities (at large $l$) and clump instabilities occurs at scales which increase with the gas fraction. This confirms that massive clumps in gas-rich discs are spatially more extended than clouds at lower \fgas. The two gas-rich cases share, once again, similarities, especially in the gas components.  The transition scale in the molecular phase is approximately independent of \fgas, which suggests that Toomre-like instabilities set the scales of the gas clumps in their atomic phase, but not in the molecular gas, confirming our findings of \sect{drivers}. Although the exact reason is difficult to identify, this could mean that collapsing or collapsed structures are dense enough to decouple from the dynamical influence of large-scale instabilities.

The regimes A', B, and C are almost inexistent in our simulations, and only found in a few measurements at the smallest scales: $l \lesssim 500 \pc$ in \htwo for A', and $l \lesssim 100 \pc$ in \hi for C, and not for the stars. These measurements are subject to numerical noise and should thus be taken with caution, but we note a tendency of gravity to become the main driver of the response of the media at sub-cloud scales, at all gas fractions. If confirmed with simulations at higher resolution, this would indicate that the gas fraction of a disc galaxy only influences the instability regime leading to the formation and assembly of the clouds/clumps but that, once the clumps are formed, the instabilities \emph{within} the clouds are always set by a dominance of gravity, with only a mere regulation from turbulence. As noted in \sect{drivers}, this picture evolves when the clumps form stars and feedback increases the turbulence within the clumps. However, at this stage of the star formation process, the gas remains coupled to stellar-driven instabilities in the gas-rich galaxies.

In the inter-clump medium, we have shown that feedback plays a more important role in the decoupling than gravitational effects (\sect{decoupling}). There, pressure stabilized the small-scales (regime D). This confirms that the change in instability regime is driven by hydrodynamics.

The existence of a transition scale of the instability regime at a few $100 \pc$ explains the differences in physical regimes noted in \sect{resolution} (between $70$ and $700 \pc$, i.e. on each sides of the transition) for the gas-rich cases: since the massive clumps have sizes of the order of 1 kpc, a given beam can have a large fraction of its area in the clump-driven regime of instability. Therefore, measuring a single value of \qrf at $700 \pc$ or an average of values measured at $70 \pc$ do not lead to the same conclusion as shown in \sect{resolution}. At low gas-fraction however, the stellar component lies in the same regime of instability at all the scales considered here. Since the stars are the main driver of instability in \gs (\fig[s]{qmaps} and \ref{fig:mpdf}), the transition seen in the gas phases has only a weak impact on the results, which explains the weak dependence of the values of \qrf with the scale, as noted in \fig{smooth}.

From the absence of a clear transition-scale in the dominant component (stars) of our \gs simulation, our results predict that the instability regime of contemporary, local disc galaxies can be correctly assessed observationally even at low resolution. The new and next generations of high resolution surveys of local discs could provide valuable insights on this aspect (e.g. \citealt{Leroy2021}). However, the transition from the disc-driven to clump-driven instability regime that we found at higher gas fraction calls for sub-100 pc resolution in high-redshift discs.

\section{Conclusions}

Using hydrodynamical simulations of isolated galaxies, we study the impact of the gas fraction on the dynamics of discs. Our main results are as follows.
\begin{itemize}
\item The scale-dependence of the structure of the ISM at large scales ($\gtrsim 100 \pc$) does not significantly depends on the gas fraction. Below this scale however, in addition to a transition to the three-dimensional regime of turbulence seen in all cases, gas-rich galaxies yield shallower power-spectra than discs at low gas-fractions. These differences arise in the molecular gas.
\item In comparison with local molecular clouds, the massive clumps of gas-rich galaxies are not significantly denser, but span larger volumes. Their star formation activity is thus not particularly more efficient, but only more productive (i.e. higher SFRs but similar depletion times).
\item Our results confirm the inadequacy of the Toomre-$Q$ criteria to assess the stability of realistic discs. They however validate the formalism of the multi-component parameter of \citet{Romeo2013}. Furthermore, this formalism allows to probe scales smaller than the disc scale-height, as now commonly done in observational surveys of nearby discs.
\item Most of the instabilities are driven by the stars, even at high gas fraction. However, the gas is the most unstable component in the inter-clump medium of gas-rich galaxies. While the formation of molecular clouds at lower gas fraction is set by stellar instabilities, the material available for the formation of massive clumps at high \fgas lies in volumes where instabilities are preferentially driven by the gas. This remains valid during the assembly phase of the clumps, until star formation starts.
\item After the onset of star formation, feedback originating from the high SFR strongly strengthens the turbulence support within massive clumps, which tends to stabilize the gaseous phase. Then, the stellar component takes over as the driver of instabilities. The stellar and gaseous instabilities in the clumps are coupled, meaning that the gas reacts to the driving from the stars. In turn, this maintains the star formation activity in the clumps, despite high levels of feedback. 
\item Decoupling between the stellar and gas phases occurs at the edge of dense regions, where feedback increases the gas velocity dispersion without significantly altering the stellar component. This process appears to be particularly efficient on the trailing side of spiral arms with high pitch angles, indicating an indirect role of the galactic-scale dynamics.
\item A too low resolution can lead to a misinterpretation of the stability regime of galaxies, in particular in gas-dominated discs. Since gas-rich disc galaxies are found at high redshift and thus observed with limited angular resolution, our results are a word of caution for the interpretation of the dynamics of clumpy galaxies inferred from observations. They also advocate against direct comparisons between studies at different spatial resolutions, for instance using gravitational lensing.
\item A transition is found in the instability regime of all galaxies, at a few $100 \pc$. At large-scale, disc dynamics sets a Toomre-like regime of instabilities, while the cloud-/clump-driven regime takes over at smaller scales. This is particularly important in the clumpy galaxies, but is mitigated at low gas fraction, where the (dominant) stellar component lies in a disc-driven regime at all scales.
\item On almost all the aspects studies in this paper, we note strong resemblances between our two gas-rich cases, while the galaxy at low gas fraction systematically stands apart. Therefore, an important change in the instability regime, of the dynamics and of the structure of disc galaxies occurs for $\fgas \approx 20\%$. This allows for the transition from clumpy turbulent discs toward spiral galaxies.
\end{itemize}

Our results, in particular on the variation of the physics regimes probed at different resolution, provides insights for the observational programs able to capture this range of scales (e.g. PHANGS-ALMA, \citealt{Leroy2021}). They also warn that these questions become even more critical in gas-rich disc galaxies as found at redshift $\approx 1\mh 3$, and that limited resolution for such distant objects could cause severe misinterpretations on the physical state of the systems.

By varying the gas fraction in our disc galaxies, we mimic the evolution of one property of galaxies across cosmic time. In principle, one could translate the gas fractions of our simulation into epochs of galaxy evolution, and even redshift, by comparing them with the fraction reported in observational surveys. However, the lack of cosmological context and thus of a self-consistent evolution of the galaxies forbids us to draw firm conclusions on the exact nature of the transition between the clump-dominated phase of galaxy evolution in gas-rich discs, and the spiral-dominated phase at lower gas fractions \citep[see also][]{Elmegreen2014}. Although our controlled experiments highlight the impact of a specific parameter on the dynamics, other aspects like the disc growth, gas accretion, and galaxy mergers would draw a more complex picture. Our conclusions thus call for further studies, both observational and theoretical, on the onset of the secular regime of disc galaxy evolution.

\section*{Acknowledgements}
We thank Bruce Elmegreen, Andreas Burkert, Angela Adamo, and Kearn Grisdale for their constructive input, and the referee for a detailed report. FR and OA acknowledge support from the Knut and Alice Wallenberg Foundation. OA acknowledges support from the Swedish Research Council (grants 2014-5791 and 2019-04659). The simulations have been performed on the Tetralith supercomputer hosted at NSC and accessed through a SNIC allocation.

\section*{Data availability}
The data underlying this article will be shared on reasonable request to the corresponding author.

\bibliographystyle{mnras}
\bibliography{biblio}

\appendix

\section{Initial conditions}
\label{sec:init}

\tab{ics} lists the initial conditions given to the \magi code to setup the models.

\begin{table}
\centering
\caption{Initial conditions. The three values corresponds to the \gs, \gm, and \gl models respectively.}
\label{tab:ics}
\begin{tabular}{|l c c c|} 
\hline
\emph{Dark matter halo} (NFW) \\ 
mass [$\times 10^{10} \Msun$] & \multicolumn{3}{c}{65.0} \\
scale length [kpc] & \multicolumn{3}{c}{28} \\
truncation radius [kpc] & \multicolumn{3}{c}{50}\\
\hline
\emph{Bulge} (Hernquist) \\
mass [$\times 10^{10} \Msun$] & \multicolumn{3}{c}{0.3} \\
scale radius [kpc] & \multicolumn{3}{c}{0.4} \\
truncation radius [kpc] & \multicolumn{3}{c}{10}\\ 
\hline
\emph{discs} (exponential) \\
total mass [$\times 10^{10} \Msun$] &  \multicolumn{3}{c}{5.5} \\
stellar mass [$\times 10^{10} \Msun$] & 4.6 & 2.6 & 1.4 \\
gas mass [$\times 10^{10} \Msun$] & 0.9 & 2.9 & 4.1 \\
scale radius [kpc] & \multicolumn{3}{c}{3.0} \\
scale height [kpc] & \multicolumn{3}{c}{0.5} \\
truncation radius [kpc] & \multicolumn{3}{c}{20}\\
\hline
\end{tabular}
\end{table}

\section{Short-wavelength and epicyclic approximations}
\label{sec:approx}

\begin{figure}
\centering
\includegraphics{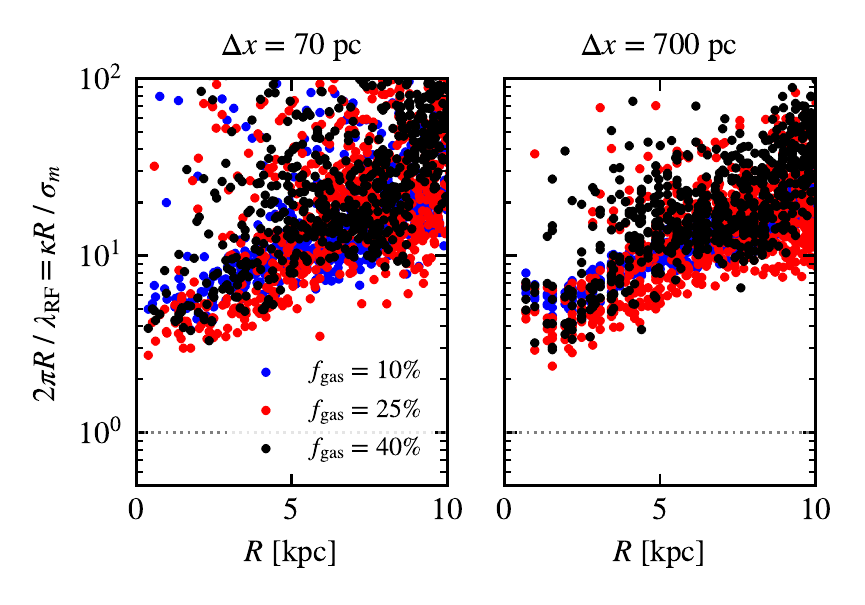}
\caption{Validation of the short-wavelength and epicyclic approximations, at high (left) and low resolution (right). In the high resolution case, only 1\% of the points are displayed, for clarity. All regions lie above unity, and more than 95\% are above 10 (see \eqn{approx}).}
\label{fig:approx}
\end{figure}

\fig{approx} demonstrates the validity of our approach on the instability analysis, by showing that the simulations fulfill the short-wavelength and the epicyclic assumptions. The former states that the inverse of the characteristic instability wavenumber should be much smaller than the galacto-centric radius, i.e. $\lrf / 2\pi \ll R$ (see \citealt{Romeo2013}). The latter imposes that the characteristic velocity dispersion is negligible with respect to the epicyclic velocity, i.e. $\sigma_m \ll \kappa R$. \eqn{lrf} shows that these two conditions are equivalent, and can written as
\begin{equation}
\label{eqn:approx}
\frac{2\pi R}{\lrf} = \frac{\kappa R}{\sigma_m} \gg 1.
\end{equation}
\fig{approx} shows that all regions from our stability analysis fulfill this condition, at both high and low resolution, and for all gas fractions: more than 99\% of the regions having a ratio greater than 3, and more than 95\% are above 10. Therefore, this validates the short-wavelength approximation made in the stability analysis.

\section{\lrf at low resolution}
\label{sec:lowres}

\begin{figure*}
\centering
\includegraphics{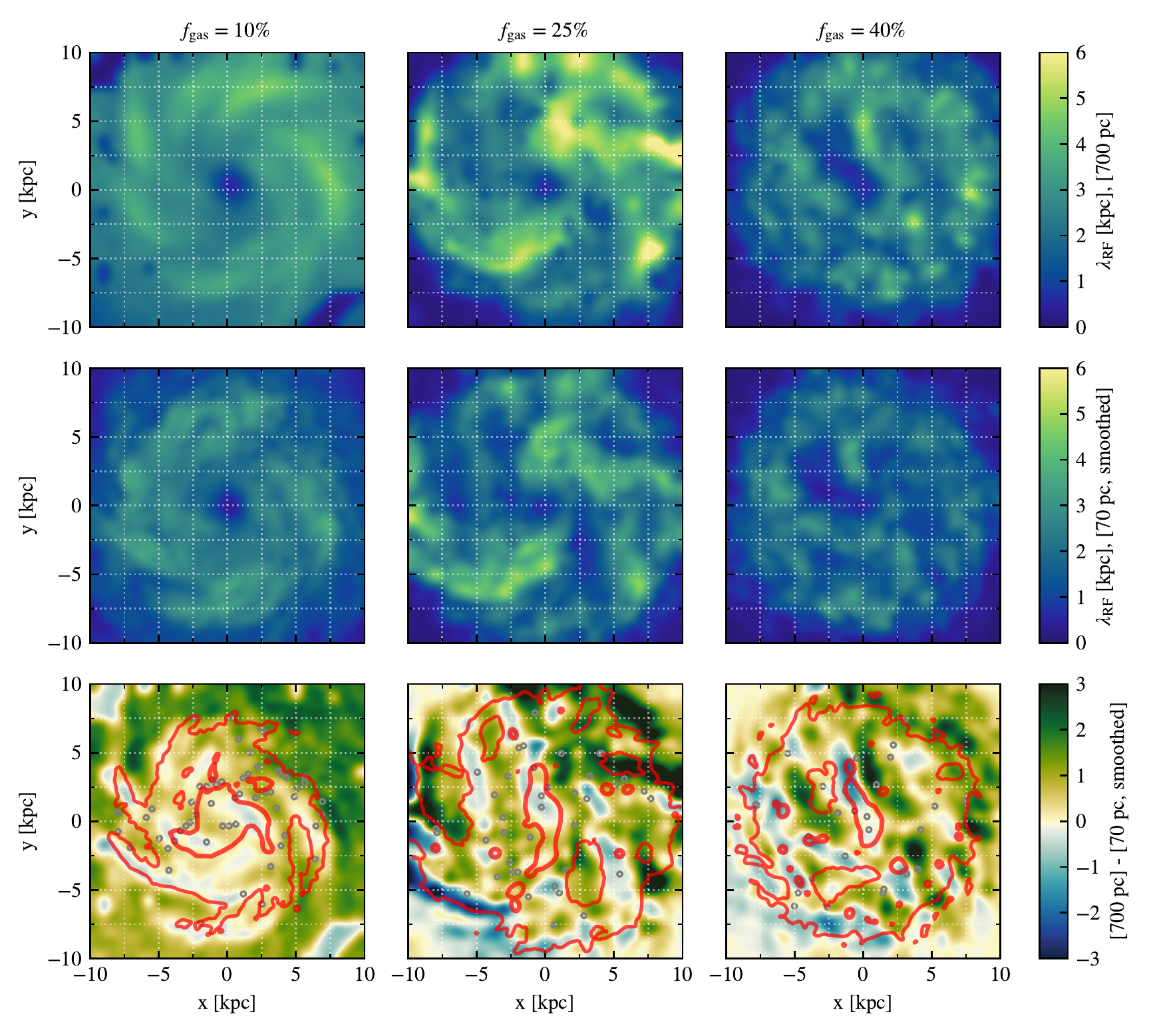}
\caption{Same as \fig{smooth}, but for \lrf.}
\label{fig:smoothl}
\end{figure*}

\fig{smoothl} shows the comparison between high and low resolution maps of the \lrf. As noted for \qrf (\fig{smooth}), the differences found here depict a change of the physical regime of instabilities between the two scales considered.

\end{document}